\documentclass[aps,prx,showpacs,notitlepage,floatfix, superscriptaddress,nofootinbib, 12pt]{revtex4-1}
\usepackage[utf8]{inputenc}
\usepackage{color,amsmath,amssymb}
\usepackage[dvipsnames]{xcolor}
\usepackage{tikz}
\usepackage{braket}
\usepackage[mmddyyyy]{datetime}
\usepackage[titletoc]{appendix}

\usepackage[colorlinks,linkcolor=blue,anchorcolor=blue,citecolor=blue,urlcolor=blue]{hyperref}
\def\be{\begin{equation}}
\def\ee{\end{equation}}
\def\bea{\begin{equation}\begin{aligned}}
\def\eea{\end{aligned}\end{equation}}

\begin{document}

\title{A Sparse Model of Quantum Holography}

\author{Shenglong Xu}
\affiliation{\footnotesize Department of Physics \& Astronomy, Texas A\&M University, College Station, Texas 77843, USA}

\author{Leonard Susskind}
\affiliation{\footnotesize Department of Physics, Stanford University, Stanford, California 94305, USA}

\author{Yuan Su}
\affiliation{\footnotesize Joint Center for Quantum Information and Computer Science, University of Maryland, College Park, Maryland 20742, and Institute for Quantum Information and Matter, California Institute of Technology, Pasadena, California 91125, USA}

\author{Brian Swingle}
\affiliation{\footnotesize Department of Physics, Brandeis University, Waltham, Massachusetts 02453, USA and Department of Physics, University of Maryland, College Park, Maryland 20742, USA}

\begin{abstract}
    We study a sparse version of the Sachdev-Ye-Kitaev (SYK) model defined on random hypergraphs constructed either by a random pruning procedure or by randomly sampling regular hypergraphs. The resulting model has a new parameter, $k$, defined as the ratio of the number of terms in the Hamiltonian to the number of degrees of freedom, with the sparse limit corresponding to the thermodynamic limit at fixed $k$. We argue that this sparse SYK model recovers the interesting global physics of ordinary SYK even when $k$ is of order unity. In particular, at low temperature the model exhibits a gravitational sector which is maximally chaotic. Our argument proceeds by constructing a path integral for the sparse model which reproduces the conventional SYK path integral plus gapped fluctuations. The sparsity of the model permits larger scale numerical calculations than previously possible, the results of which are consistent with the path integral analysis. Additionally, we show that the sparsity of the model considerably reduces the cost of quantum simulation algorithms. This makes the sparse SYK model the most efficient currently known route to simulate a holographic model of quantum gravity. We also define and study a sparse supersymmetric SYK model, with similar conclusions to the non-supersymmetric case. Looking forward, we argue that the class of models considered here constitute an interesting and relatively unexplored sparse frontier in quantum many-body physics. 
\end{abstract}

\maketitle

\newpage

\tableofcontents

\section{Introduction}

The phenomenon of quantum information scrambling is of considerable interest to multiple communities for the light it sheds on issues ranging from the black hole information problem (e.g.~\cite{Hayden2007, Sekino2008,Shenker2014}) to quantum transport (e.g.~\cite{PhysRevLett.117.091601}). Scrambling is a process whereby initially simple information becomes so thoroughly mixed across the degrees of freedom of a system that it cannot be read out by any few-body measurement~\cite{Hayden2007, Sekino2008,brown2012scrambling, Shenker2014, Hosur2016}. A key quantity is the scrambling time, defined as the timescale for information to spread across the entire system. In the absence of special phenomenon like localization, a system's scrambling time is primarily determined by its geometry.

To illustrate this point, consider a system whose Hamiltonian consists of nearest neighbor couplings as defined by a finite Euclidean lattice with $N$ sites in $d$ spatial dimensions. Suppose for simplicity that the local dynamics is controlled by a single characteristic timescale $\tau$. Typically a disturbance on one site will propagate ballistically, affecting $\sim\left(t/\tau\right)^d$ degrees of freedom after time $t$. Given this growth, information will spread over the whole system in a time of order $t_* \sim \tau N^{1/d}$. Hence, the dimension of space is the key determinant of the system size dependence of the scrambling time.

In the limit of large dimension, the number of nearest neighbors diverges and an extrapolation of the $N^{1/d}$ behavior suggests a scrambling time of order $\ln N$. If the diverging number of neighbors is modeled by a system with all-to-all interactions, then many studies have indeed shown that such models, such as the Sachdev-Ye-Kitaev (SYK) model~\cite{Sachdev1992,kitaev2015,Sachdev2015,Polchinski2016,Maldacena2016}  and Gurau-Witten tensor models~\cite{Gurau2017,Witten2019},  can scramble in time $t_* \sim \tau \ln N$~\cite{Sekino2008,Lashkari2012}.  However, we emphasize that these results are not straightforward: the couplings must be normalized in a carefully chosen $N$-dependent fashion and the scrambling time is only manifest after a non-trivial calculation.

It is therefore natural to ask if a $\ln N$ scrambling time can occur in a model where each degree of freedom has a bounded number of neighbors and the system-size scaling is obvious from the geometry. A nearest neighbor Hamiltonian defined on a uniform tiling of hyperbolic space provides a simple affirmative to this question. The reason is that the volume of a sphere in hyperbolic space grows exponentially with the radius. Hence, if information is moving ballistically such that it spreads over a sphere of radius $t/\tau$ after time $t$, then the volume of that sphere can equal the total number of sites after a time of order $t_* \sim \tau \ln N$.

The same kind of estimate can also be obtained from a simple tree model, since the number of sites in a tree grows exponentially with the depth of the tree, but the distance between any two leaves is proportional to the depth and hence to $\ln N$.\footnote{We should note that the diameter of a graph is not the only relevant feature. For example, trees scramble slowly in other senses due to entanglement bottlenecks, see e.g.~\cite{harrow2019separation}.} All of the models above can be subsumed into a general setting in which we consider nearest neighbor Hamiltonians defined on arbitrary graphs of bounded degree. Among such graphs, the behavior seen in Euclidean space turns out to be very non-generic whereas the hyperbolic and tree behaviors are more generic.

For example, if we randomly choose a connected graph on $N$ vertices where each vertex has fixed bounded degree, then such a graph turns out to be locally like the tree model with high probability as $N$ grows large. This means that the tree/hyperbolic scrambling behavior is generic at large $N$. Moreover, the scrambling time is proportional to $\ln N$ for a physically transparent reason: the maximum distance between any two points is of order $\ln N$. Such models can be called sparse fast scramblers since they feature a small number of connections (compared to all-to-all models) while still scrambling in time $\sim \ln N$. 

The purpose of this paper is to generalize these intuitions by studying a sparse version of the SYK model defined using a bounded degree hypergraph. For the same physically transparent reasons as above, the system-size dependence of the scrambling time will generically be $\ln N$. Moreover, we establish that the low temperature limit of the system features the same kind of emergent gravitational dynamics as in the fully connected SYK model. This implies that our sparse model is maximally chaotic in a thermal state with large inverse temperature $\beta$ in the sense of saturating the MSS bound~\cite{Maldacena2016b},
\be 
t_* \geq \frac{\beta}{2\pi}  \log{N}.
\label{fastest}
\ee

We note that a class of generalized SYK models~\cite{Gu2017} have recently been considered on general graphs~\citep{Bentsen2019}. This class of models is defined by placing SYK clusters with many fermions and all-to-all interactions on each vertex of a graph and by letting the clusters interact according to the edge structure of the graph. By contrast, in the model we consider, each vertex of a hypergraph is associated with a single Majorana fermion, and each such fermion interacts with a bounded number of other fermions. This means that the total number of terms in the Hamiltonian scales linearly with the number of vertices. Some bounds on the dynamics of such models have recently been derived~\cite{chen2019operator}.

There are a number of reasons why the kind of sparse model we consider is interesting. On a conceptual level, the existence of a wider class of fast scramblers exhibiting maximal chaos at low temperature opens the door to a broader class of systems that might exhibit black hole behavior. Moreover, because the system size dependence of the scrambling time is transparent in sparse models, they may serve as effective theories of fast scrambling dynamics for densely connected systems. On a practical level, sparse systems typically admit much more efficient computer simulations---both classical and quantum. By significantly reducing the resources needed to simulate black holes in holographic models of quantum gravity, these results bring us closer to the goal of studying ``quantum gravity in the lab'' in the sense of ~\cite{brown2019quantum}. Such experiments are likely still some years away, but one recent proposal showed how sparse interactions (in a simpler spin model) could be generated with cold atoms coupled to a cavity~\cite{Bentsen2019a}. Previous work specifically considering quantum simulations of SYK include~\cite{Danshita_2017,PhysRevX.7.031006,PhysRevLett.119.040501,Babbush2019,cao2020quantum}.

Here, we are primarily interested in fast scrambling physics and in the SYK model as a toy model of low-dimensional quantum black holes. As such, our focus is on determining whether the gravitational physics persists in the sparse model. Based on a path integral analysis and on numerical results with up to 52 fermions, our conclusion is that sparse SYK does fast scramble at all temperatures and exhibits a maximally chaotic gravitational sector at low temperature provided $k$ is large enough. We do not determine the minimal value of $k$ required for gravitational physics to emerge at low temperature, but our numerical data suggest $k_{\min} < 4$ when $q=4$. There is a simple lower bound $k_{\min} > 1/q$.

More generally, the random constructions we utilize here can be used to construct families of sparse models from a wide variety of fully connected models. This represents a potentially rich class of models that can be studied both analytically and with quantum simulators and computers. Here we show that this sparse frontier includes new models of quantum gravity; further explorations may shed light on issues ranging from efficient quantum error correcting codes to the robustness of entanglement at finite temperature.

The remainder of this paper is organized as follows. Section~\ref{sctn:general} sets the stage with a review of some relevant graph theory and a discussion of the basic timescales for scrambling in lattice versus all-to-all models. Section~\ref{sctn:ssyk} contains the definition and path integral analysis of sparse SYK. Section~\ref{sctn:susy_ssyk} does the same for sparse supersymmetric SYK. Section~\ref{sctn:smltn} reports the results of numerical simulations with up to 52 fermions and analyzes the quantum Hamiltonian simulation cost of the sparse models. Section~\ref{sctn:discussion}
contains a discussion of the results and future directions.

\section{General quantum many-body physics on expander graphs}
\label{sctn:general}

Consider a quantum many-body system with $N$ degrees of freedom, which can be spins, electrons, etc. Each degree of freedom lives on a vertex. In its most general form, the Hamiltonian is a sum of $L$ terms
\bea
H= \sum \limits_{i}^L H_{\{v_i\}},
\eea
where $H_{\{v_i\}}$ acts on a set of vertices denoted as $\{v_i\}$. The Hamiltonian defines a hypergraph. Each term in the Hamiltonian represents a hyper-edge connecting a set of vertices $v_{i}$. Hamiltonians which are sums of terms acting on $q$ vertices called $q$-local\footnote{The term $q$-local is often used for Hamiltonians built from terms with $q$ vertices or less. Here we will use it when all terms act on exactly $q$ vertices.}. The corresponding hypergraph is called $q$-uniform. Some useful properties and terminology about graphs have been collected in Appendix~\ref{sctn:graphs}.

\subsection{2-local Hamiltonians on graphs}
\label{sctn:2-local}

Let us first consider 2-local systems defined on graphs with bounded degree. In some respects, these systems are similar to conventional lattice models representing discretizations of continuous space. The interactions are between nearest neighbors, of which there are a finite number, and they feature Heisenberg-like couplings. For example, at high temperature, short-time local properties are insensitive to the global structure of the graph, so we may expect all generic 2-local models to behave similarly. However, since most graphs do not have translation symmetry, a defining feature of a lattice, 2-local Hamiltonians defined on generic graphs are expected to behave differently at longer times or lower temperatures. 

One property of interest is the time scale $t_*$ for information scrambling. Scrambling is naturally tied to the growth in complexity of initially simple Heisenberg operators. As time increases, a Heisenberg operator, say $\sigma^z_v(t)$, that originally only has support on vertex $v$, gradually spreads to more vertices. Correspondingly, the initial information about the operator---which operator it was and where it started---becomes impossible to recover locally. The scrambling time $t_*$ is roughly the time-scale for the operator to cover all vertices.

The following picture is helpful to give a rough estimate of $t_*$. Intuitively, operator growth is like distributing some secret information. Imagine a graph populated by people who are going to play the telephone game. One person has a secret to distribute. In the first round of play, that person tells all their nearest neighbors (on the graph) the secret. In the next round, whomever knows the secret tells all their nearest neighbors. How long does it take before everyone knows the secret? Let $S_t$ be the set of vertices that know the secret at time $t$. The timescale can estimated from a simple continuum model,
\bea
\frac{d |S_t|}{dt} =\lambda |\partial S_t|,
\label{eq: epid}
\eea 
where $|S_t|$ is the number of the vertices that already know the secret and $|\partial S_t |$ is the number of neighboring vertices. The dynamics stops and the system is scrambled when $|S_t| \sim N$, the total number of vertices. We can estimate the scrambling time $t_*$ if we know the relation between $|S|$ and $|\partial S|$ for sphere-like regions $S$. If the graph is an ordinary $n$ dimensional lattice, $|\partial S| \sim |S|^{1-1/n}$ and we get $t_*\sim N^{1/n}.$ 

Now lattices are very sparse when interpreted as graphs since the number of edges grows linearly with $N$. An obvious way to speed up scrambling is to increase the number of terms in the Hamiltonian, e.g., making the interactions all-to-all. This works, but the resources required, e.g., the number of interaction terms, increase accordingly. Depending on the goal, this increase may be acceptable, but as we will see below, it does increase the cost of simulating the system using a quantum computer.

How to make a fast scrambler with minimal resources, i.e., a minimal set of interaction terms in the Hamiltonian? One has to look for systems beyond lattice models. Ideally, we want the number of terms in the Hamiltonian to scale the same way as in lattice models, linearly with the number of vertices. At least linear scaling of edges with vertices is certainly necessary for fast scrambling since sub-linear scaling leads to disconnected components.

There is a class of sparse graphs called expanders that possess much stronger connectivity than lattices. Given a subgraph $S$ of an expander $G$, as long as its size is smaller than half the total size, the number of neighboring vertices outside the subgraph is always proportional to the size. In other words, $|\partial S|\sim |S|$, $\forall |S|< |G|/2  $. Expander graphs, thought of as networks, are both sparse and very efficient at distributing information. Plugging the relation into Eq.~\eqref{eq: epid} immediately gives rise to fast scrambling behavior $t_*\sim \log N$. This is the reason that systems defined on expanders can be fast scramblers~\cite{Bentsen2019}. The difference between lattices and expanders can also be understood from the diameter of graph, defined as the length of the path between the two furthest separated vertices. The diameter of a lattice scales as $D \sim N^{1/n}$ whereas $D\sim \log N$ for an expander. 

In graph theory, it is well-known that the minimal ratio between $|\partial S| $ and $|S|$ for $|S|<|G|/2$, called Cheeger number $h(G)$, is bounded by the spectral gap $\Delta$ of the Laplacian matrix of the graph $G$~\cite{cheeger},
\bea
2 h(G)\geq \Delta \geq \frac{h^2 (G)}{2 d _\text{max}},
\eea
where $d$ is the degree associated with each vertex. As a result, a finite spectral gap implies finite Cheeger number and vice versa. Based on the Cheeger inequality, it is straightforward to identify expanders from the spectral gap of the Laplacian for sparse graphs. We mention this because graph Laplacians will feature prominently in our analysis in Section~\ref{sctn:ssyk}.

\begin{figure}
\includegraphics[height=0.9\columnwidth, width=0.9\columnwidth]
{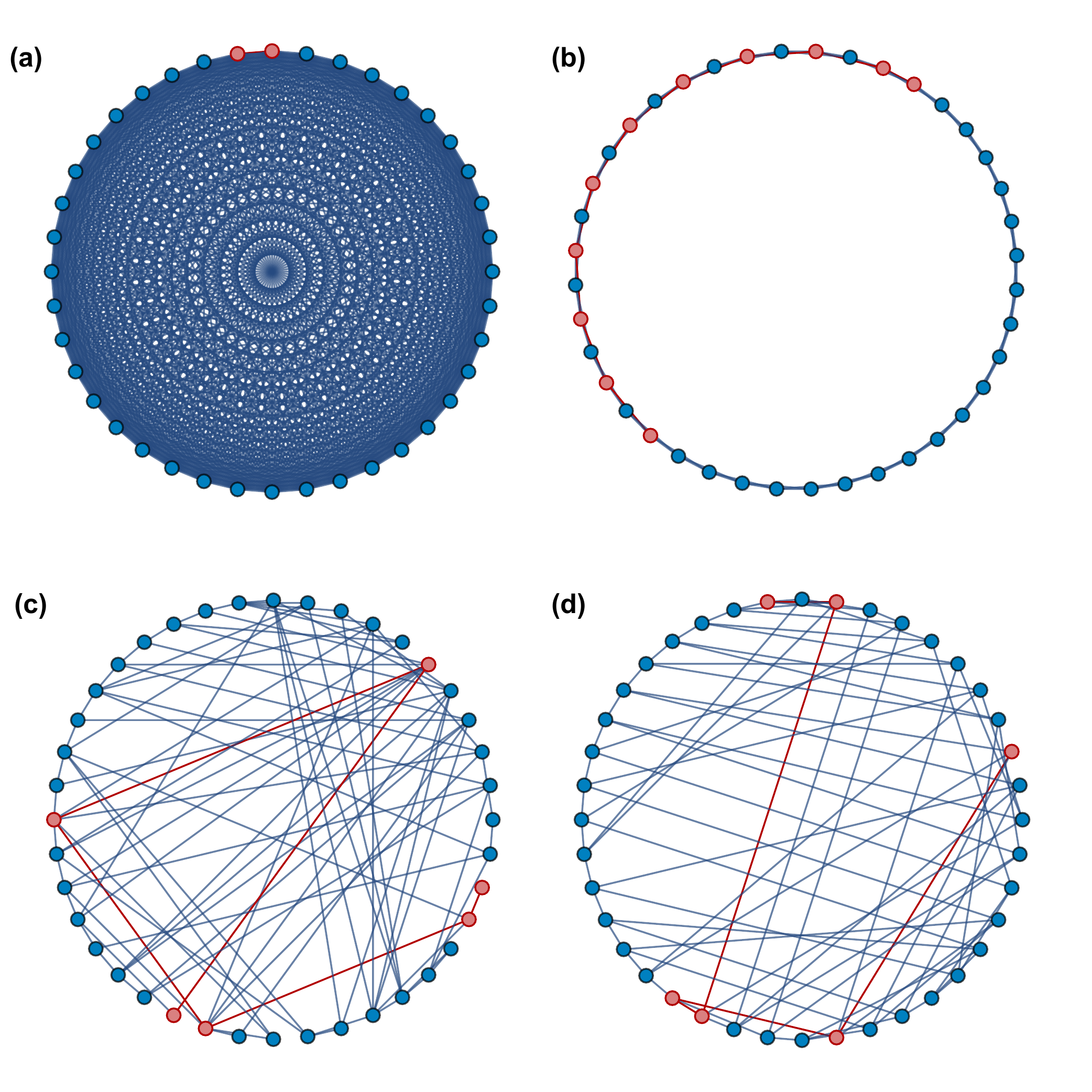}
\caption{Comparison between complete graph, lattice, random graph and random regular graph. The diameters of the graphs are highlighted. }
\label{fig:graph}
\end{figure}

Are graphs with these nice properties difficult to find? It turns out that expanders are rather generic while lattices are special. In fact random graphs with fixed degree $d$ for every vertex, called random regular graphs, are known rigorously to be expanders. Such graphs have finite spectral gap $\Delta \sim d-2\sqrt{d-1}$~\cite{Alon1986,Friedman2003}. In addition, the number of edges, and hence the number of terms in the Hamiltonian $L$, is $d N/2$ which is linear in $N$. One can also relax the regular condition and consider sparse random graphs. Sparse random graphs are generated by  pruning edges independently with the edge probability $p=2k/N$, so that the average number of edges are $k N$ and the degrees satisfy the Poisson distribution $P(2k)$. Based on results of Erd\"os and R\'enyi~\cite{Erdos1960}, when $k>1/2$, the graph breaks up into a number of disconnected components with one \it giant component \rm containing of order $N$ vertices and with all other components being small of size $\log{N}$. We may throw away the disconnected components and retain only the giant component. This giant component is believed to be an expander, at least for large enough $k$, because of its similarity to the random regular graph. In Fig. \ref{fig:graph}, we provide examples of (a) a complete graph, (b) a one-dimensional lattice, (c) a random graph, and (d) a random regular graph, along with the path between the two vertices which are furthest apart.

\subsection{$q$-local Hamiltonians on hypergraphs}
Now we turn to $q$-local Hamiltonians ($q>2$), corresponding to hypergraphs. One reason to consider cases beyond $2$-local is that $2$-local fermionic models are non-interacting and therefore do not scramble in the many-body Hilbert space. For definiteness, we will consider a system of $N$ real fermionic degrees of freedom $\psi_{\alpha} $
with $\alpha = 1,2,.....,N$. The Hamiltonian is a sum of operators of the form $\psi_{\alpha_1}\psi_{\alpha_2}...\psi_{\alpha_q}$.

Based on the simple information spreading picture in Eq. \ref{eq: epid}, we may estimate the system's scrambling behavior by considering the interaction graph obtained from the underlying hypergraph determined by the Hamiltonian. The interaction graph is defined by including an edge between two vertices if and only if $H$ contains an interaction term that includes both degrees of freedom. Note that the relevant Hamiltonian terms will also include other degrees of freedom. Vertices connected by edges are, by definition, nearest neighbors on the graph. Thus, by definition, the Hamiltonian is nearest neighbor coupled on the graph. If the interaction graph is an expander, then it is reasonable to expect that the Hamiltonian exhibits fast scrambling. In the following, based on the discussion in Sec. \ref{sctn:2-local}, we describe two classes of hypergraphs that are expected to yield interaction graphs that are expanders. We also note that there are numerous proposals to directly characterize the expansion of hypergraphs without necessarily referring to the interaction graph (for example, see~\cite{Lubotzky2018} and references therein), but for our purposes below, the interaction graph turns out to be the key object. 

\textit{Random Graph}---The first class consists of random hypergraphs obtained from random pruning. In a $q$-uniform hypergraph with $N$ vertices, there are maximally $\binom{N}{q} \approx N^q/q!$ hyperedges. Starting from the complete hypergraph, we delete each hyperedge independently with probability $1-p$ where $p=kq!/N^{q-1}$ so that there are $kN$ remaining hyper-edges on average. Then given two vertices, the probability that there is at least one hyper-edge connecting them is,
\bea
1 - (1-p)^{\frac{(N-2)^{q-2}}{(q-2)!}} \approx \frac{k q(q-1)}{N}.
\eea  
In another words, the edge survival probability for the underlying interaction graph is $k q(q-1)/N$, which has the right scaling for a expander. Even though the edge probabilities are not independent as is the case for random pruning of graphs, we still expect that the resulting interaction graph is close to a random graph and therefore is an expander. In addition, with $k$ terms in the Hamiltonian, the edge probability is amplified by the factor $q(q-1)/2$ compared to that of 2-local Hamiltonians, and thus we expect that the expansion property enhances as $q$ increases while the number of terms in the Hamiltonian stays the same. For example, the connected cluster with order $N$ vertices appear for $k>1/(q(q-1))$ instead of $1/2$. 

\textit{Random Regular Graph} -- Another class of $q$-uniform hypergraphs with nicer properties is random regular hypergraphs, where each vertex appears in exactly $d$ hyper-edges. A procedure for generating random regular hypergraphs can be found in an appendix. In a $d$-regular, $q$-uniform hypergraph, the total number of terms is $d N /q$, so we set $d=kq$ to keep the number of total terms equal to $kN$. Then a vertex interacts with $d(q-1)$ other vertices on the condition that a pair of vertices does not appear together in more than one hyper-edge, the probability of which is suppressed by $1/N^2$. Therefore, the underlying interaction graph is very close to a random regular graph with degree $k q(q-1)$ and is also expected to be an expander. 

The discussion in this section is intended to motivate studying expanders in the context of quantum dynamics, so the analysis only gives a rough estimation of the scrambling time of a generic many-body Hamiltonian defined on such a (hyper)graph. Many details have been ignored so far. For example, the classical description Eq.~\ref{eq: epid} only becomes accurate for 2-local random circuit models in the large spin limit. Furthermore, we have not yet discussed the temperature dependence of scrambling. In the following sections, by constructing and studying specific toy models, namely the sparse SYK models, we will establish a more precise connection between the scrambling behavior of many-body system and the underlying hypergraph structure, both regular and irregular.

\section{Sparse SYK}
\label{sctn:ssyk}
To fix notation, we first briefly recall the all-to-all SYK model with $N$ fermions and $q$-fermion interactions~\cite{Maldacena2016}. The number of interaction terms is 
\begin{equation}
   L=\binom{N}{q} \approx \frac{N^q}{q!},
\end{equation}
with the last statement valid for $N$ large with $q$ fixed. The variance of the couplings is 
\begin{equation}
    \langle J_{a_1 \cdots a_q}^2\rangle_J = \frac{(q-1)! J^2}{N^{q-1}},
\end{equation}
and the full Hamiltonian is
\begin{equation}
    H_{\text{full}}=i^{q/2}\sum_{a_1 <  \cdots < a_q} J_{a_1\cdots a_q} \chi_{a_1} \cdots \chi_{a_q} = \frac{i^{q/2}}{q!} \sum_{a_1 \neq \cdots \neq a_q} J_{a_1\cdots a_q} \chi_{a_1} \cdots \chi_{a_q}.
\end{equation}
This defines an ensemble of Hamiltonians depending on the dense set of couplings $J_{a_1 \cdots a_q}$. We use $\langle \cdots \rangle_J$ to denote an average over the ensemble of $J$ couplings. We assume familiarity with standard SYK techniques and results, see Ref.~\cite{Maldacena2016} for background.

\subsection{Sparsity from random pruning}
\label{sctn:random_pruning}
An ensemble of sparse models is defined by the following procedure. For every term in the full Hamiltonian, set it to zero with probability $1-p$, leading to a random hypergraph. The average number of terms (hyperedges in the hypergraph) remaining is 
\begin{equation}
    L= p \binom{N}{q} \approx p \frac{N^q}{q!}.
\end{equation}

The strength of the couplings should be changed to account for the loss of terms so that the physical timescales are comparable. The variance of the couplings should be set to 
\begin{equation}
\langle J_{a_1 \cdots a_q}^2\rangle_J = \frac{(q-1)! J^2}{N^{q-1}}\frac{1}{p}.
\end{equation}

There are various ways to understand this choice. For example, to leading order in a high temperature expansion, one can verify that this for the coupling variance leads to an average energy which is independent of $p$. To extend this reasoning to lower temperatures, one can appeal to a Feynman diagram analysis. Readers unfamiliar with this approach can consult Ref.~\cite{Maldacena2016}.

\begin{figure}
\begin{tikzpicture} 
  \draw[thick] (1,0) -- (5,0);
  \draw[thick] (3,0) circle [radius=1]; 
  \draw (.75,0) node {$a$};
  \draw (5.25,0) node {$a$};
  \draw (3,1.25) node {$b$};
  \draw (3,.25) node {$c$};
  \draw (3,-.75) node {$d$};
  \draw[thick,dashed] (2,0) to[bend left] (2.5,1.75) to[bend left] (3.5,1.75) to[bend left] (4,0);
\end{tikzpicture}
\\
\begin{tikzpicture} 
  \draw[thick] (1,0) -- (5,0);
  \draw[thick] (3,0) circle [radius=1]; 
  \draw (.75,0) node {$a$};
  \draw (3,1.25) node {$b$};
  \draw (3,.25) node {$c$};
  \draw (3,-.75) node {$d$};
  \draw[thick,dashed] (2,0) to[bend left] (2.5,1.75) to[bend left] (3.5,1.75) to[bend left] (4,0);
  \begin{scope}[shift={(4,0)}]
  \draw[thick] (1,0) -- (5,0);
  \draw[thick] (3,0) circle [radius=1]; 
  \draw (5.25,0) node {$a$};
  \draw (3,1.25) node {$b'$};
  \draw (3,.25) node {$c'$};
  \draw (3,-.75) node {$d'$};
  \draw[thick,dashed] (2,0) to[bend left] (2.5,1.75) to[bend left] (3.5,1.75) to[bend left] (4,0);
  \end{scope}
\end{tikzpicture}
\caption{Basic and iterated melon diagram with $q=4$.} \label{fig:melon}
\end{figure}
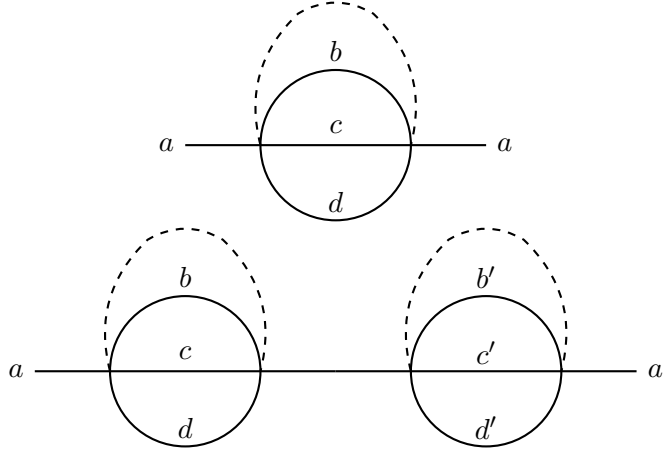

Consider the basic melon diagram in Figure~\ref{fig:melon}. With $p=1$, so that every term survives, the melon diagram gives a factor of
\begin{equation}
    \frac{1}{2} \times \frac{1}{(q!)^2} \times \frac{(q-1)! J^2}{N^q} \times 2 \times q \times q! \times N^q = J^2.
\end{equation}
Here is the explanation for the factors: the $1/2$ from second order perturbation theory, the $1/q!$ in the Hamiltonian, the coupling variance, the number of interaction vertices to connect the incoming line to, the number of ways to connect that incoming line to the vertex, the number of ways to connect the first vertex to the second, and the number of internal configurations summed over.

Now, what happens when $p \neq 1$? Two things change. First, on average, the number of non-vanishing internal configurations (configurations for which $J_{a_1 \cdots a_q} \neq 0$ after the pruning process) is reduced by a factor of $p$. Second, the variance of the couplings is modified. These two factors precisely cancel, ensuring that the melon diagram gives the same contribution.

Next consider the iterated melon shown in Figure~\ref{fig:melon} with $p\neq 1$. Naively, the same argument as for the single melon applies. Indeed, for all the terms for which $b'c'd'$ is not a permutation of $bcd$, the single melon analysis applies. However, for those terms for which $b'c'd'$ is a permutation of $bcd$ (a fraction $3!/N^3$ of all terms, or $(q-1)!/N^{q-1}$ for general $q$), the $J$ couplings vanish or not all together. The result is that this term contributes
\begin{equation}
    p \times \frac{(q-1)!}{N^{q-1}} \frac{1}{p^2} = \frac{(q-1)!}{p N^{q-1}}.
\end{equation}
In fact, there are three such terms, corresponding to the three ways to Wick contract the $J$ couplings (normally these terms would be $1/N$ suppressed). Hence, this special subset of configurations contributes
\begin{equation}
    3 \frac{(q-1)!}{p N^{q-1}} J^4.
\end{equation}

From this expression, one key expansion parameter is  $\frac{(q-1)!}{p N^{q-1}}$. We may understand this as follows. Suppose there is an average of $k N$ terms remaining after the pruning,
\begin{equation}
    L = k N.
\end{equation} 
The pruning probability $p$ is thus
\begin{equation}
    p = \frac{k N}{\binom{N}{q}} \approx \frac{k q!}{N^{q-1}}.
\end{equation}
The quantity $k$ is roughly analogous to the degree of a graph. Hence, if $k$ grows with $N$, then the number of terms in the Hamiltonian grows faster than linearly with $N$ and, to leading order at large $N$, the fully connected physics is recovered. 

In terms of $k$, the expansion parameter identified above is
\begin{equation}
    \delta = \frac{(q-1)!}{p N^{q-1}} \approx \frac{1}{kq}.
\end{equation}
The appearance of $q$ makes sense, since even with only $k N$ terms, a single fermion will typically appear $k q $ times due to the fact that it has $q$ chances to appear in each term.

Below we will analyze the sparse model using a path integral approach, but we note that one can also formulate a diagram analysis in which one expands in $1/N$ and $1/k$. To order $k^0$, the usual set of melon diagrams familiar from the fully connected model also give the sparse model physics. However, at large $N$ and finite $k$, there are corrections organized in a $1/k$ expansion.

\subsection{Sparsity from regular hypergraphs}

The pruning procedure introduced above has the virtue of being straightforward to implement. However, while it is statistically regular, there are sample to sample fluctuations in, for example, the number of terms in the Hamiltonian and the connectivity of each fermion. At large $N$, pruning also generically gives some disconnected clusters of fermions that do not interact with the bulk of the system.

These issues may be circumvented by defining sparse SYK models using regular hypergraphs. Recall that a hypergraph is just a collection of vertices and edges (or hyperedges) in which each edge may have more than two vertices in it. A hypergraph is called $d$-regular if every vertex is in $d$ edges and it is called $q$-uniform if every edge contains exactly $q$ vertices.

To define a sparse SYK model with $kN$ terms in the Hamiltonian, we need a $q$-uniform hypergraph with $N$ vertices which is $kq$-regular. The resulting model is specified by giving $x_{a_1 \cdots a_q} \in \{0,1\}$ which denote which interactions are present. The regularity condition implies that
\begin{equation}
    \frac{1}{(q-1)!}\sum_{a_2 \cdots a_q} x_{a_1 a_2 \cdots a_q} = k q
    \label{eq:regular_graph}
\end{equation}
independent of $a_1$.

The Hamiltonian is then
\begin{equation}
    H=\frac{i^{q/2}}{q!} \sum_{a_1 \neq \cdots \neq a_q} J_{a_1\cdots a_q} x_{a_1 \cdots a_q} \chi_{a_1} \cdots \chi_{a_q}.
\end{equation}
where the couplings have variance
\begin{equation}
    \langle J_{a_1 \cdots a_q}^2 \rangle_J = \frac{J^2}{kq}.
\end{equation}
Of course, this form of the Hamiltonian can also describe the randomly pruned case provided the $x_{a_1 \cdots a_q}$ are taken to be independent random variables with $\Pr(x_{a_1\cdots a_q} =1)=p$.

\subsection{Disorder averaged partition function}

It is convenient to analyze the sparse model using a path integral approach. Consider first the path integral for a single realization of the disorder, 
\bea
    Z = \int D\chi e^{- I_E}
\eea
where the action on the Euclidean contour is
\bea
    I_E = \int d\tau \left( \frac{1}{2}\sum_a \chi_a \partial_\tau \chi_a + H \right).
\eea
The physically interesting quantity is the disorder average of the logarithm of $Z$. In the case of fully connected SYK, the partition function turns out to be self-averaging provided the temperature is not too low, so one may first compute the disorder averaged partition function and then take its logarithm. However, in the sparse case, the partition function is no longer self-averaging, so we must take the logarithm and then compute the disorder average. We emphasize that this statement has nothing to do any glass physics, but is simply a consequence of considering an ensemble of Hamiltonians where the number of terms is proportional to the number of degrees of freedom. 

We use the replica trick to access the disorder average of $\ln Z$. Recall that this is accomplished by representing $\ln Z$ as the $m \rightarrow 0$ limit of $\frac{Z^m-1}{m}$. The goal is to compute the disorder average of $Z^m$ as a function of $n$ and then take the limit $m \rightarrow 0$. We discuss in detail the sample-to-sample fluctuations of various physical quantities in the appendix~\ref{sctn:flctn}.

Before any averaging, the partition function of $m$ copies with identical disorder is
\begin{equation}
    Z^m = \int \prod_\alpha D\chi^\alpha \exp\left( -\int d\tau\left[ \frac{1}{2} \sum_{a,\alpha} \chi_a^\alpha \partial_\tau \chi_a^\alpha + i^{q/2} \sum_{A,\alpha} J_A x_A \chi_{a_1}^\alpha \cdots \chi_{a_q}^\alpha \right] \right).
\end{equation}
Here $\alpha=1,\cdots,m$ is a replica index. 
Next, we average over the couplings $J_A$ with $x_A$ fixed to obtain the action,
\bea
    I_E= \frac{1}{2} \int d\tau \sum_{a,\alpha} \chi_a^\alpha \partial_\tau \chi_a^\alpha - \frac{\braket{J_A}^2}{2} \sum_{A,\alpha,\beta} x_A \Phi_A^{\alpha,\beta},
\eea
where
\begin{equation}
    \Phi_A^{\alpha,\beta} = \int d\tau d\tau' \chi^\alpha_{a_1}(\tau) \chi^\beta_{a_1}(\tau') \cdots \chi^\alpha_{a_q}(\tau) \chi^\beta_{a_q}(\tau').
\end{equation}
It is convenient to introduce the fields
\begin{equation}
    G_{a}^{\alpha \beta}(\tau_1,\tau_2) = \chi_a^\alpha(\tau_1)\chi_a^\beta(\tau_2)
\end{equation}
and the corresponding Lagrange multipliers $\Sigma^{\alpha \beta}_a$. Then the action becomes
\bea
    I_E&=\frac{1}{2}  \int d\tau \sum_{a,\alpha} \chi_a^\alpha \partial_\tau \chi_a^\alpha -\frac{1}{2} \int d\tau \sum _{a,\alpha, \beta} \chi_a^\alpha (\tau_1)\chi_a^\beta (\tau_2)\Sigma_a^{\alpha\beta} \\ &+\frac{1}{2}\int d\tau_1 d\tau_2 \left (\sum _{a,\alpha, \beta} \Sigma_a^{\alpha\beta}G_a^{\alpha\beta}-\braket{J_A^2}\sum_{A,\alpha,\beta} x_A G_{a_1}^{\alpha\beta}...G_{a_q}^{\alpha\beta} \right ).
\eea
After integrating out the $\chi$ fields, we obtain the action for $\Sigma$ and $G$,
\bea 
I_E =& -\frac{1}{2} \sum\limits_a \log \text{det}\left(-i\omega \delta^{\alpha\beta} -\Sigma_a^{\alpha\beta}(\omega)\right)\\ 
+ &\frac{1}{2}\int d\tau_1 d\tau_2 \left (\sum _{a,\alpha, \beta} \Sigma_a^{\alpha\beta}G_a^{\alpha\beta}-\braket{J_A^2}\sum_{A,\alpha,\beta} x_A G_{a_1}^{\alpha\beta}...G_{a_q}^{\alpha\beta} \right ).
\label{eq:ssyk_single_graph_action}
\eea 

\subsection{Fixed interaction graph}
\subsubsection{Random regular hypergraph}
\label{sctn:regular}
The equations of motion (EOM) for a fixed set of $x_A$ are
\bea
      \partial_{\tau_1} G^{\alpha\gamma}_a(\tau_1, \tau_2) - \int d\tau_2 G^{\alpha\beta}_a (\tau_1, \tau_2) \Sigma^{\beta\gamma}_a (\tau_2, \tau_3) =\delta^{\alpha\gamma} \delta(\tau_1-\tau_3)   \\
        \Sigma_a^{\alpha \beta}= \braket{J^2_A} \sum_{a_2 <\cdots< a_q} x_{a, a_2\cdots a_q} G_{a_2}^{\alpha \beta} \cdots G_{a_q}^{\alpha \beta}.
     \label{eq:syk_sp_eom}
\eea 
For this section, we postulate that the interactions arise from regular hypergraph with uniform degree $d=kq$ as discussed above. In this case, the number of terms in the summation is $kq$. Based on Eq. \ref{eq:regular_graph}, there exists a uniform solution independent of the fermion label $a$. In terms of the replica indices, we assume the dominant saddle point is replica diagonal. 

Before proceeding with this ansatz, it should be noted that there can be replica symmetric but non-diagonal solutions or even replica symmetry breaking solutions. In the fully connected case, non-diagonal saddle points do exist, but these saddles never dominate the path integral which computes the free energy, at least not for any reasonable temperature~\cite{Arefeva2018}. Now, in the sparse case arising from a regular hypergraph, the saddle point action of a uniform (independent of fermion label) saddle point is identical to the fully connected saddle point action. Hence, from the fully connected results, the replica diagonal saddle point always dominates, at least among uniform saddle points. This analysis applies only the to saddle point action, but in the sparse case, fluctuations also contribute $O(N)$ to the free energy. Hence, in principle one should minimize the full free energy, meaning the saddle point action contribution plus the fluctuations, over different saddle points. However, because the fluctuation contribution is suppressed by $1/k$ while the difference between different uniform saddle point actions is independent of $k$, it follows that the saddle point action alone still determines the dominant saddle point provided $k$ is large enough. Of course, this does not rule out non-uniform saddles of various types; if these exist, they may be relevant at low temperatures. 

Returning to the diagonal and uniform ansatz, the equations of motion imply that the diagonal elements obey the same Schwinger-Dyson equation as in the fully connected model provided the coupling variance is $\braket{J_A^2}_J =J^2/(kq)$. Note that this is precisely the value we previously identified using a diagram analysis. Hence, independent of the hypergraph degree, we have
\begin{equation}
    G_s^{-1} = - i \omega - \Sigma_s, \ \ \ \Sigma_s = J^2 G^{q-1}_s.
\end{equation}
However, if the degree is too small, then the physics does change since the hypergraph breaks up into pieces. For example, a regular hypergraph with $d=1$ is just $N/q$ disconnected all-to-all graph with $q$ fermions. We can also construct disconnected regular hypergraphs even for higher $d$. These properties of the hypergraph do not change the symmetric solution but play an important role in the fluctuation on top of the mean-field solution, which we discuss in detail below.

Next we consider quadratic fluctuations around this saddle point. The action for the quadratic fluctuations has three pieces. The first one comes from the second derivatives of the Pfaffian term, the second piece comes from the Lagrange multipliers, and the last piece comes from the interaction term. The first two pieces are the same as the ones in the all-to-all SYK model, while the third piece depends on the underlying hypergraph structure. 

Following the standard approach in~\cite{Maldacena2016}, we obtain the action of the quadratic fluctuation as 
\bea
\delta I_E &= \frac{1}{4}\sum\limits_{a,\alpha\beta}\int d \tau  G(\tau_1, \tau_3)G(\tau_2, \tau_4)\sigma_a^{\alpha\beta}(\tau_1, \tau_2)\sigma_a^{\alpha\beta}(\tau_3, \tau_4)\\
&+\frac{1}{2} \sum\limits_{a, \alpha\beta}\int d\tau   \sigma_a ^{\alpha\beta}(\tau_1, \tau_2) g_a^{\alpha\beta}(\tau_1, \tau_2)\\
&-\frac{1}{4} J^2 (q-1) \int d\tau \sum\limits_{ab,\alpha\beta} M_{ab} G_s^{q-2}(\tau_1, \tau_2) g_a^{\alpha\beta}(\tau_1, \tau_2)  g_b^{\alpha\beta}(\tau_1, \tau_2)
\eea 
where $g$ and $\sigma$ stand for the fluctuation of the green function and the self-energy respectively, and the adjacency matrix $M_{ab}$ is defined as
\bea
M_{ab} = \frac{1}{kq (q-1)}\sum\limits_{a,b, a_3<\cdots<a_q} x_{a b a_3\cdots a_q}.
\eea

Because we expand the action around a replica-diagonal saddle, only the replica-diagonal components of the interaction have fluctuation at the quadratic order. The only nonzero source of replica non-diagonal fluctuations at the quadratic order are the fermion determinants since the interaction only contributes to the action for replica off-diagonal fluctuations at order $q$. Furthermore, at quadratic order, the diagonal fluctuations completely decouple from the off-diagonal fluctuations. Integrating out the $\sigma$ fields and rescaling the $g$ fields following~\cite{Maldacena2016}, we obtain a quadratic action $\delta I_E = m \delta I_{E,d} + m(m-1) \delta I_{E, nd}$. The replica diagonal part $I_{E, d}$ and the off-diagonal part $I_{E, nd}$ are
\bea 
 &\delta I_{E,d} =\frac{J^2 (q-1)}{4}\sum\limits_i \tilde{g}_i (K^{-1}-\lambda_i) \tilde{g}_i + \frac{N}{2}\log \text{det} K, \quad  \tilde{g}_i =\sum\limits_a g_a u_{ai}\\
 &\delta I_{E, nd} = \frac{J^2 (q-1)}{4}\sum g_a K^{-1} g_a + \frac{N}{2}\log \text{det} K.
\eea
where $\lambda_i$ and $u$ are the eigenvalues and normalized eigenvectors of the adjacency matrix $M$, and $K$ is the usual kernel appearing in the fully connected SYK model. The term $\frac{N}{2}\log \text{det} K$ comes from integrating the $\sigma$ fields. This term makes sure that $I_{nd}$ is zero at the quadratic order after integrating out the $g$ fields.

At this point, the analysis is general to any temperature. Moreover, at very high temperature, one can verify by explicit computation that, to leading order in $\beta$, the average energy is identical between the sparse and dense models. However, this explicitly fails at the next non-trivial order, as shown in Appendix~\ref{sctn:high temperature}, so the sparse and dense models do not have identical thermodynamics. Nevertheless, as we now investigate, they may have qualitatively similar low temperature properties in the regime where gravity emerges in the dense model. We now analyze the quadratic fluctuations in the sparse model at low temperature. 

The spectrum of the kernel $K^{-1}$ is well studied at low temperature based on an emergent conformal symmetry (Eq. 3.73 in \cite{Maldacena2016}). It exhibits a set of discrete positive eigenvalues starting at $1$ and a continuum of negative eigenvalues below a finite negative value. As a result, all the modes in the replica off-diagonal part $\delta I_{E, nd}$ are gapped and independent of the interaction graph. On the other hand, the diagonal part does depend on the adjacency matrix of the hypergraph. Remember that we are so far considering the case where the hypergraph is regular. As a result, $\sum\limits_a M_{ab}=1$, namely the largest eigenvalue of $M$ is 1, corresponding to the uniform mode on the graph, $\tilde{g}_0=\frac{1}{\sqrt{N}}\sum\limits_a g_a$. Separating the uniform mode from the rest, we can write the diagonal quadratic fluctuation as,
\bea
\delta I_{E,d} = \frac{J^2 (q-1)}{4}\tilde{g}_0 (K^{-1}-1) \tilde{g}_0 + \frac{J^2 (q-1)}{4}\sum\limits_i \tilde{g}_i (K^{-1}-\lambda_i )\tilde{g}_i + \frac{N}{2}\log \text{det} K.
\label{eq:quadratic_diagonal}
\eea 

So far, the analysis is valid for any regular hypergraph. Let us first go back to the all-to-all case where the underlying hypergraph is both regular and complete. In this case, all the non-identity eigenvalues of $M$ are equal to $1/N$.  Because of the spectrum of $K$ in the conformal limit, only the first term in Eq. \ref{eq:quadratic_diagonal} contributes a soft mode, meaning a mode whose fluctuations are not suppressed in the conformal limit. This soft mode is expected to control the low-energy dynamics since all the non-uniform modes are gapped. The soft mode is the temporal reparameterization mode and gives rise to the Schwarzian derivative when perturbed away from the conformal limit. This is the usual story in the all-to-all SYK model. 

In the sparse limit, where the number of interaction terms scales linearly with $N$, the non-identity eigenvalues of $M$ are not concentrated with $1/N$ of zero but occupy a band of some non-vanishing width at large $N$. If the spectrum of $K^{-1}$ overlaps with the eigenvalues of any of the non-uniform modes of $M$, then there will be additional soft modes that will presumably significantly alter the SYK physics. Therefore it is important to analyze the spectrum of $M$ in more detail.

The matrix $M$ is the generalized adjacency matrix of the $kq$-regular $q$-uniform hypergraph. Clearly one must have $k>1/q$ and with $kq$ an integer. This matrix $M$ can be generated from the incidence matrix $A$ of size $N \times k N$. The element $A_{ij}=1$ if the $i$th vertex is in the $j$th hyperedge, and is $0$ otherwise. One can show that the adjacency matrix $M$ of the hypergraph can be obtained by summing over the hyperedge indices, 
\bea 
M= \frac{1}{kq (q-1)} \left ( AA^T - kq I \right ).
\label{eq:M_reg}
\eea
The positivity of the matrix $A A^T$ guarantees that the spectrum of $M$ is bounded below by $-1/(q-1)$ for any regular hypergraph. We note that the incidence matrix can be interpreted as the biadjacency matrix for a bi-regular graph. A more refined analysis of the spectrum of the corresponding bi-regular graph for random regular hypergraph~\cite{Brito2018, Dumitriu2019} reveals that the eigenvalues of $M$, except the largest ones, fall into the following range,
\bea
\max (0, \sqrt{kq-1}-\sqrt{q-1})^2-kq  \leq kq  (q-1) \lambda \leq 2\sqrt{kq-1}\sqrt{q-1} +q -2.  
\label{eq:spectrum_adj}
\eea
The range reduces to that of random regular graph~\cite{Alon1986, Friedman2003} when $q=2$.

\begin{figure}
\includegraphics[height=0.45\columnwidth, width=0.45\columnwidth]
{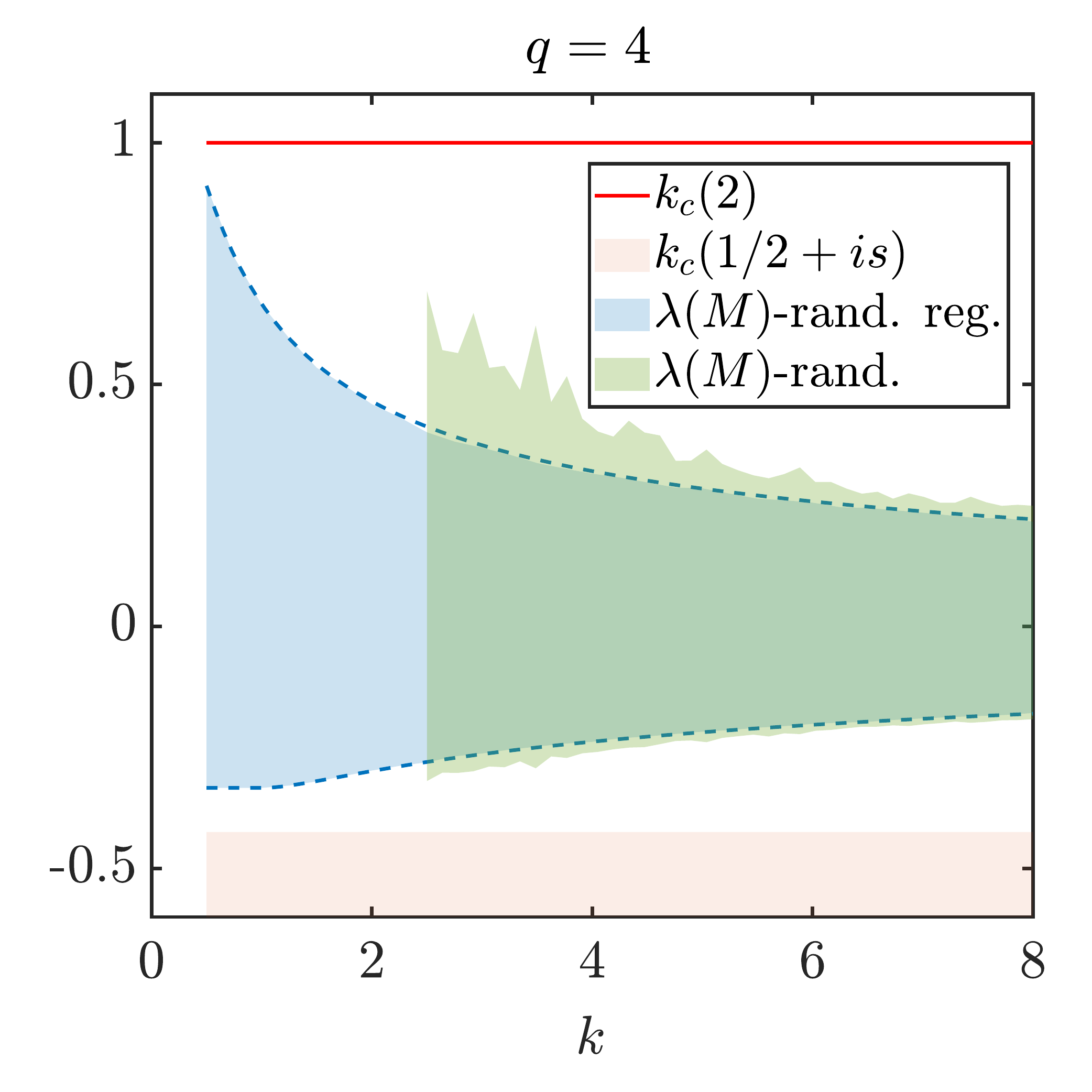}
\includegraphics[height=0.45\columnwidth, width=0.45\columnwidth]
{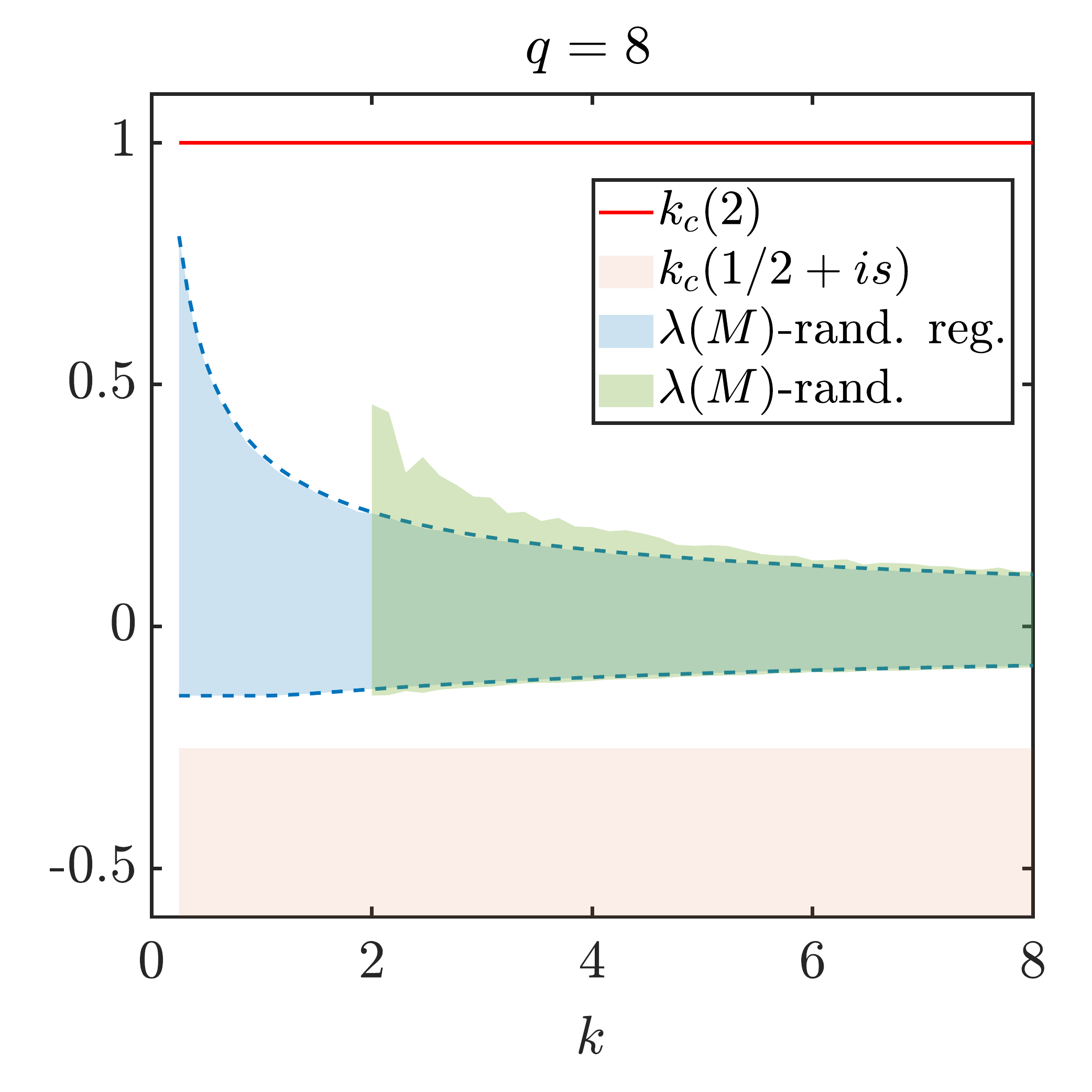}
\caption{Comparison between the spectrum of the SYK kernel and the adjacency matrix of the  hypergraph $M_{ab}$. The shaded blue region represents the spectrum range (excluding the uniform mode) of $M$ in Eq.~\ref{eq:M_reg} obtained from random regular hypergraph numerically, which agrees with the analytical upper and lower limits quoted in Eq.~\ref{eq:spectrum_adj}. The shaded green region represents the spectrum range of $M$ in Eq.~\ref{eq:M_gen} obtained from random hypergraph. In both cases, a typical graph with 1000 vertices are considered at each $k$. The gap between the lower limit and the spectrum of the SYK kernel $k_c(1/2)$ is non-zero for $M$ from any hypergraph, while the gap between the upper limit and $k_c(2)=1$, also known as the spectrum gap of the graph, is non zero for $M$ of a typical hypergraph. The two gaps ensures that the only soft mode of the sparse SYK model defined on a typical graph is the same as the all-to-all SYK model. On the other hand, when the sparse SYK model is defined on disconnected graph or lattices, the spectrum gap of the adjacency matrix vanishes and there are additional soft modes contributing to the low energy physics. }
\label{fig:spectrum}
\end{figure}

The resulting interplay between the spectrum of $M$ for randomly generated regular hypergraphs and the SYK kernel $K^{-1}$ is presented in Fig.~\ref{fig:spectrum}. In particular, from the numerically obtained spectrum of $M$, we see that the results are consistent with the above bounds and that the spectrum fits neatly between the discrete modes and continuum of $K^{-1}$. fitting into the gap between the discrete modes and the continuum of $K^{-1}$. 
The top of the continuum of the SYK model is $k_c^{-1}(1/2)$, which is always smaller than the lower bound $-1/(q-1)$ of $M$ for any regular hypergraph. Similarly, the gap between the uniform mode and the rest of the spectrum of $M$ is finite for a typical regular random graph. The two finite gaps indicate that, for typical interaction graph, the quadratic fluctuations of all non-uniform modes are gapped, and the only enhanced contribution comes from the uniform mode, similar to the case of the all-to-all SYK model.

Of course, there exist regular hypergraphs where the spectrum of $M$ is gapless near the unit eigenvalue. The simplest example is when the hypergraph contains disconnected subgraphs, in which case each subgraph contributes one mode with eigenvalue 1. Graphs with translation symmetry (lattices) are another example. These additional soft modes will also contribute to the low energy physics and are left for future study.

To summarize, we have shown that there exists a uniform saddle point solution of the equations of motion which is identical to the fully connected solution. In the low temperature limit, we also showed that the quadratic fluctuations around this saddle point are gapped except for the uniform mode, whose quadratic action is identical to the action in the fully connected case. However, we emphasize that these two facts alone do not imply that the thermodynamics, say, of the sparse model is identical to that of the fully connected model. We know explicitly from the high temperature expansion that they are not identical. Moreover, from the quadratic action, we see that integrating over the non-uniform modes contributes a term $\sum\limits_i \log \text{Det} (1-K \lambda_i)$, which scales as $N/(kq)$ to the free energy. This term should be computable.

What these results strongly suggest is that the low temperature state of the sparse model exhibits the same pattern of symmetry breaking as the fully connected model. In particular, there should be an emergent conformal symmetry at low energy which is both spontaneously and weakly explicitly broken~\cite{Maldacena2016,maldacena2016conformal}. The Goldstone mode of this spontaneously broken symmetry is precisely the gravitational mode~\cite{Maldacena2016,maldacena2016conformal}, so one expects to recover a gravitational sector at low energy, albeit with possibly renormalized parameters. Hence, we conjecture for $k > k_\text{min}$, the sparse model has a gravitational sector exhibiting maximal chaos at low temperature.

\subsubsection{Random hypergraph}

In this section we go back to the equation of motion in Eq.~\ref{eq:syk_sp_eom} and consider generic hypergraphs which are not necessarily regular. For example, when a hypergraph is generated from random pruning as described in Sec.~\ref{sctn:random_pruning}, the degrees of all vertices satisfy a Poisson distribution $P(kq)$. The averaged degree is $kq$ and the standard deviation is $\sqrt{kq}$. Therefore a random hypergraph is not regular for general $k$. As a consequence, Eq.~\ref{eq:syk_sp_eom} does not admit a uniform solution. 

Solving the equation of motion for general hypgraph graph at arbitrary coupling strength becomes intractable even numerically because there are $2N$ instead of 2 coupled integral-differential equations. Fortunately, at low energy, the equations can be solved by exploiting the usual reparameterization symmetry since the argument in the all-to-all case is still valid. When the non-interacting propagator, the $-i\omega$ term in EOM, is ignored, the time dependence of the Green functions is completely set by the conformal symmetry, and only the coefficients can vary based on the hypergraph structure. Following the convention in~\cite{Maldacena2016}, we get
\bea
G_{a}(\tau)=\gamma_a \frac{b}{|\tau|^{2\Delta}}\text{sgn}(\tau)=\gamma_a G(\tau), \ \ \ \Sigma_a = \frac{J^2}{kq}\sum\limits_{a a_2< \cdots < a_q }x_{a a_2\cdots a_q } \gamma_2\cdots \gamma_q G(\tau)^{q-1}=\frac{1}{\gamma_a}\Sigma(\tau)
\eea 
Different from the all-to-call case, there are site-dependent factors $\gamma_i$, satisfying 
\bea
\frac{1}{kq}\sum \limits_{a_2<\cdots <a_q }x_{a_1 a_2\cdots a_q} \gamma_{a_1}\gamma_{a_2} \cdots \gamma_{a_q}=1
\label{eq:gamma}
\eea
which depend only on the hypergraph structure. Intuitively, these parameters $\gamma_a$ compensate the fluctuations in the degree and regularize the graph. It remains an open question whether the solution exists for any hypergraph or only for some specific type of hypergraph. We will discuss this issue later and assume the solution exists for now. 

Assuming the solution exists, we are ready to analyze the quadratic fluctuation around this saddle point. The off-diagonal part decouples and is gapped, just as we for regular hypergraphs,. Considering only the replica diagonal part for simplicity, the quadratic action is
\bea
\delta I_{E, d} 
= \frac{1}{4}\int \gamma_a^2 G^s(\tau_1, \tau_3)G^s(\tau_2, \tau_4) \sigma_a (\tau_1, \tau_2)\sigma_a (\tau_3, \tau_4) - \frac{1}{2}\int d\tau g (\tau_1, \tau_2)  \sigma(\tau_1, \tau_2) \\
- \frac{1}{4}J^2(q-1)\int d\tau \sum \limits_{ab} M_{ab}G_s^{q-2}(\tau_1, \tau_2) \gamma_{a}^{-1} \gamma_{b}^{-1}  g_a (\tau_1,\tau_2) g_b (\tau_1,\tau_2),
\eea
where the symmetric matrix $M$ is defined as
\bea
M_{ab}=\frac{1}{kq (q-1)}\sum\limits_{a_3<\cdots <a_q}x_{ab a_3\cdots a_q} \gamma_a \gamma_b \gamma_{a_3}\cdots \gamma_{a_q}.
\eea
We rescale the fields $g_a \rightarrow \gamma_a G_s^{1-q/2} g_a $ and $\sigma_a \rightarrow \gamma_a G_s^{q/2-1} \sigma_a $ and integrate out the $\sigma$ fields to obtain the action for $g$,
\bea
\delta I_{E, d} = \frac{J^2(q-1)}{4}  \left ( g_a K^{-1}  g_a -  g_a M_{ab}  g_b\right),
\eea
where $K$ again is the kernel for the all-to-all SYK model. Here $M$ plays the same role as the adjacency matrix in the regular case studied in Sec.~\ref{sctn:regular}, but it is sensitive to the $\gamma_a$. Because of Eq. \ref{eq:gamma}, $M_{ab}$ has a uniform eigenvector with eigenvalue 1, giving rise to a soft mode in the action when combined with the $k_c(2)$ mode from the SYK kernel $K^{-1}$. This is the only soft mode if $M$ is gapped, making the low energy physics similar to that of all-to-all SYK model. 

Again we see that the low energy properties of the sparse SYK model with a fixed graph reduce are controlled by the spectral properties of $M$. Similar to the case of regular hypergraphs, $M$ can be generated by an incidence matrix $A$,
\bea
M = \frac{1}{kq (q-1)} (A\Gamma A^T - kq I )
\label{eq:M_gen}
\eea
The matrix $\Gamma$ is diagonal, with $\Gamma_j =\gamma_{i_1}\gamma_{i_2}\cdots \gamma_{i_q}$, where $i_1\cdots i_q$ are the vertices in the $j$th hyperedge. Because $A\Gamma A^T $ is positive, the spectrum of $M$ is also always bounded blow by $1/(q-1)$ and does not overlap with the continuum in $K^{-1}$ for any hypergraph. On the other hand, determining an upper bound on the spectrum for a typical random hypegraph is challenging due to these unknown coefficients $\gamma$. Therefore we turn to a numerical solution of Eq. \ref{eq:gamma} and construct $M$ for a random hypergraph to obtain the range of the spectrum. 

Empirically we find that the solution $\gamma_i$ is close to uniform for large $k$ and fluctuates severely vertex by vertex for small $k$. This is expected as the degrees of a random hypergraph graph vary significantly at small $k$. In addition, searching for a solution becomes difficult for the numerical solver when $k$ is below $\sim 2$. In Fig. \ref{fig:spectrum}, we include the spectrum range of $M$ for random hypergraph generated by pruning with $1000$ vertices for both $q=4$ and $q=8$ and $k$ down to where the solver can find the solution efficiently. We find that the upper limit of $M$ (excluding the unit eigenvalue) is above that of the regular graph for small $k$ but approaches the regular case as $k$ increases. This agrees with the expectation that a random hypergraph becomes statistically regular in the large $k$ limit. 

From this analysis we conclude that even the randomly pruned sparse model likely exhibits a gravitational sector for sufficiently large $k$. In principle, the minimal value of $k$ could be different in the pruned case, due to the greater degree of fluctation in the local hypergraph properties.

\subsection{Averaging over prunings}
\label{sctn:ssyk_pruning}

In the last subsections, we discussed the sparse SYK model defined on a fixed hypergraph. Here we consider the sparse model where we also average over random prunings. Let us go back to Eq. \ref{eq:ssyk_single_graph_action}. Averaging over random prunings can be done for each factor in the interaction part of the partition function as follows,
\bea
\left\langle  \exp \left (\frac{J^2}{2kq} \sum_{\alpha\beta}\int d\tau_1 d\tau_2 x_A G^{\alpha\beta}_{a_1}\cdots G^{\alpha\beta}_{a_q} \right)\right\rangle_x &=1+p \exp \left (\frac{J^2}{2kq}\sum_{\alpha\beta} \int d\tau_1 d\tau_2  G^{\alpha\beta}_{a_1}\cdots G^{\alpha\beta}_{a_q} \right)-p \\
&\approx  \exp\left \{ p \exp \left (\frac{J^2}{2kq}\sum_{\alpha\beta} \int d\tau_1 d\tau_2 G^{\alpha\beta}_{a_1}\cdots G^{\alpha\beta}_{a_q} \right)-p \right \}.
\eea
The second line is valid for $p \sim 1/N^{q-1}$; we only consider this case since if $p$ decreases more slowly with $N$, then the pruned model is manifestly close to the fully connected model at large $N$. 

With this formula for the average over prunings, the replicated action becomes
\bea 
I^{(n)}_E =& -\frac{1}{2} \sum\limits_a \log \text{det}\left(-i\omega \delta^{\alpha\beta} -\Sigma_a^{\alpha\beta}(\omega)\right)\\ 
&+\frac{1}{2}\int d\tau_1 d\tau_2  \sum _{a,\alpha, \beta} \Sigma_a^{\alpha\beta}G_a^{\alpha\beta}- \sum\limits_A 
2p \exp \left (\sum\limits_{\alpha\beta}\frac{J^2}{2kq} \int d\tau_1 d\tau_2 G^{\alpha\beta}_{a_1}\cdots G^{\alpha\beta}_{a_q} \right).
\eea 
As a result, the second equation of motion in Eq. \ref{eq:syk_sp_eom} is modified to
\bea
\Sigma_a^{\alpha\beta} = \frac{(q-1)!}{N^{q-1}} J^2\sum\limits_{a_2 \cdots a_q \neq a} \exp \left (\sum\limits_{\alpha'\beta'}\frac{J^2}{2kq} \int d\tau_1 d\tau_2 G^{\alpha'\beta'}_{a} G^{\alpha'\beta'}_{a_2}\cdots G^{\alpha'\beta'}_{a_q} \right)G^{\alpha\beta}_{a_2}\cdots G^{\alpha\beta}_{a_q}.
\eea
It still admits a uniform replica-diagonal solution satisfying
\bea
\Sigma_s= J^2 \exp\left(\frac{mJ^2}{2kq}\int d\tau_1 d\tau_2  G_s^q\right)G_s^{q-1},
\eea
where $m$ is the number of the replicas.  This is the same as the EOM in the all-to-all SYK model but with $J^2$ enhanced by a extra factor.  Notice that the exponent goes to zero in the replica limit $m \rightarrow 0$. In the zero replica limit, the equation of motion is the same as in fully connected SYK and $J^2$ is not renormalized.

Hence, at the level of the saddle point action and non-including fluctuations, the uniform replical-diagonal solution gives exactly the same free energy as the fully connected model. Since we know that the free energy receives $1/k$ corrections, this cannot be the whole story. Of course, there are the usual caveats about the possibility of other saddles being dominant. However, there is another complication: the number of distinct fields being integrated over is proportional to $N$, since the action is not just a function of $\sum_a G_a^{\alpha \beta}$. Therefore, the fluctuations around the saddle point can in principle make a contribution of order $N$ to the free energy. Hence, as with a fixed interaction graph, the correct procedure is to find saddle points, compute the full free energy including the saddle point action and fluctuations, and minimize the free energy to determine the equilibrium state.

We have not carried out the fluctuation analysis for this formulation. The full path integral does does enjoy a permutation symmetry among the fermions, which should strongly constrain the fluctuations. However, the action is also more non-local in time, since it contains arbitrary powers of integrals over time (from the exponential term in the action). So this is an interesting problem for future work.

\section{Sparse supersymmetric SYK}
\label{sctn:susy_ssyk}
\subsection{Disorder averaged partition function}

In this section, we construct a sparse version of the $\mathcal{N}=1$ supersymmetric (SUSY) SYK model originally introduced in \cite{Fu2016}. The starting point is the formula for the supercharge $Q$. The all-to-all construction is
\bea
    Q = i^{\frac{q-1}{2}} \sum_{a_1<a_2<\cdots <a_q} C_{a_1 a_2 \cdots a_q} \chi_{a_1} \chi_{a_2}...\chi_{a_q}.
\eea
where $q$ is an odd number and the couplings $C_{abc}$ are Gaussian random variables with mean zero and variance
\bea
    \langle C_{a_1\cdots a_q}^2 \rangle_C = \frac{(q-1)!}{N^{q-1}}J.
\eea
The $C$ couplings are totally antisymmetric. Note that due to the prefactor, the superchange $Q$ is Hermitian. The Hamiltonian $H$ is simply $Q^2$ and thus positive.  

In the sparse version, we keep only a subset of the terms contributing to $Q$. For example, with random pruning each term is kept with probability $p$, so that the sparse supercharge is
\bea
    Q = i ^\frac{q-1}{2}\sum_{a_1<\cdots <a_q} C_{a_1\cdots a_q} x_{a_1\cdots a_q} \chi_{a_1} \cdots \chi_{a_q}
\eea
where for each term, $x_{a_1\cdots a_q} = 1$ with probability $p$ and $x_{a_1\cdots a_q} = 0$ with probability $1-p$. To compensate the reduction in the number of terms, the normalization of the variance must be modified. We determine the this modification below.

Note that throughout this section $p$ refers to the probability that a term in the supercharge survives pruning. Similarly, $k$ is defined to be the ratio of the number of terms in the supercharge to the number of fermions. Hence, these parameters have a different meaning in this section than in the preceding section on the non-SUSY model. In particular, while the number of terms in the SUSY Hamiltonian is still proportional to $N$ in the sparse limit, the ratio of the number of Hamiltonian terms to number of fermions is no longer $k$.

As discussed in \cite{Fu2016}, it is simplest to analyze this model by introducing a set of auxiliary boson degrees of freedom $\phi_a$. The replicated Euclidean path integral is
\begin{equation}
    Z = \int D\chi D\phi e^{ -I_E}
\end{equation}
where
\bea 
I_E =\sum\limits_{\alpha=1}^m \int d\tau \left (\frac{1}{2}\chi_a^\alpha \partial
   _{\tau }\chi_a^\alpha - \frac{1}{2}\phi _a^\alpha \phi
   _a^\alpha +i^{\frac{q-1}{2}}\sum
   _{a, a _2<\cdots <a _q} C_{a  a _2  \cdots a_q}\phi _a^\alpha \chi _{a_2}^\alpha \cdots \chi _{a _q}^\alpha \right)
\eea

Averaging the partition function over the couplings $C$ gives a new effective action, 
\bea
  I_E =& \int_0^\beta d\tau \sum_{a,\alpha} \left(\frac{1}{2}\chi_a^\alpha \partial_\tau \chi_a^\alpha - \frac{1}{2} \phi_a^\alpha \phi_a^\alpha \right)  \\
  &- \frac{\braket{C^2}}{2}\int d\tau_1 d\tau_2 \sum_{\alpha\beta, a_1, a_2<\cdots <a_q} x_{a_1 a_2\cdots a_q} \phi_{a_1}^\alpha (\tau_1)\phi_{a_1}^\beta (\tau_2) \chi_{a_2}^\alpha (\tau_1) \chi_{a_2}^\beta (\tau_2)\cdots \chi_{a_q}^\alpha (\tau_1)\chi_{a_q}^\beta (\tau_2)  \\
  &+ \frac{\braket{C^2}}{2}\int d\tau_1 d\tau_2 \sum_{\alpha \beta,a_3<\cdots <a_q} x_{a_1 \cdots a_q} \phi^\alpha_{a_1}(\tau_1) \chi^\beta_{a_1}(\tau_2) \chi^\alpha_{a_2}(\tau_1)\phi^\beta_{a_2}(\tau_2)  \chi^\alpha_{a_3}(\tau_1) \chi^\beta_{a_3}(\tau_2)\cdots \chi_{a_q}^\alpha (\tau_1) \chi_{a_q}^\beta (\tau_2).
\eea
The terms have been naturally grouped into different correlation functions, either $\chi \chi$, $\phi \phi$, or $\chi \phi$. The standard approach is to introduce the Lagrange multipliers $\Sigma^{\chi \chi, \phi\phi, \chi\phi, \phi\chi}$ and the Green's functions $G^{\chi \chi, \phi\phi, \chi\phi, \phi\chi}$ such that the action becomes quadratic in the fermion fields $chi$ which can then be integrated out. However, the supersymmetry of the system is not manifest in such an approach. 

To fully use the supersymmetry, one can introduce the fermionic superfield $\Phi_a(\tau, \theta)=\chi_a(\tau) +\theta \phi_a(\tau)$ where $\theta$ is a Grassman number, and write the action in a manifestly supersymmetric invariant manner,
\bea
I_E=
&-\frac{1}{2}\sum\limits_{a, \alpha}\int d \tau  d \theta  \Phi_a^\alpha  D_{\theta }\Phi_a^\alpha\\ &+\frac{\braket{C^2}}{2}\sum\limits_{\alpha\beta, a_1<\cdots a_q } \int d \tau
   _1 d \tau_2 d \theta_1 d \theta_2 x_A \Phi^\alpha _{a_1} (\tau _1 )\Phi^\beta 
   _{a _1}(\tau _2)\Phi^\alpha _{a _2}(\tau _1)\Phi^\beta 
   _{a _2}(\tau _2\text{)...}\Phi^\alpha _{a _q}(\tau
   _1)\Phi^\beta _{a_q}(\tau _2)
\eea 
where $D_{\theta}$ is the superderivative $\theta \partial_\tau +\partial_\theta $. 
Now, as with the non-SUSY version, we introduce a field $\mathcal{G}$ for each $a$, $\mathcal{G}_a^{\alpha \beta} (1,2)=\Phi^\alpha_a(1)\Phi^\beta_a(2)$, and corresponding Lagrange multiplier fields $\varSigma$. Note that $\mathcal{G}$ and $\varSigma$ depends on $\tau_1$ and $\tau_2$ as well as $\theta_1$ and $\theta_2$, the latter being Grassman numbers. With these new variables, the effective action becomes quadratic in the fermionic fields,
\bea
I_E=& -\frac{1}{2}\sum\limits_{a, \alpha} \int d \tau  d \theta  \Phi_a^\alpha D_{\theta }\Phi_a^\alpha + \frac{1}{2} \sum\limits_{a, \alpha\beta} \int d \tau_1 d \tau_2 d \theta_1 d \theta_2 \varSigma^{\alpha\beta}_a(1,2) \left(\Phi^\alpha_a(1)\Phi^\beta_a(2) -\mathcal{G}^{\alpha\beta}_a(1,2) \right ) \\
&+\frac{\braket{C^2}}{2}\sum\limits_{\alpha\beta, a_1<\cdots< a_q} \int d \tau
   _1 d \tau_2 d \theta_1 d \theta_2 x_A \mathcal{G}^{\alpha\beta}_{a_1}(1,2)\cdots \mathcal{G}^{\alpha\beta}_{a_q}(1,2).
  \label{eq:susy_sp_actn}
\eea
 
\subsubsection{Fixed interaction graph}
The action Eq. \ref{eq:susy_sp_actn} gives rise to the following manifestly SUSY invariant equations of motion  for a fixed hypergraph defining the supercharge,
\bea
\varSigma_a^{\alpha\beta}(1,2) = \braket{C^2} \sum\limits_{a, a_2 < \cdots <a_q} x_{a a_2\cdots a_q} \mathcal{G}^{\alpha\beta}_{a_2} (1,2) \cdots \mathcal{G}^{\alpha\beta}_{a_q} (1,2)\\
D_{\theta_3} \mathcal{G}_a^{\alpha\gamma} (1,3) +\int d \tau_2 d \theta_2 \mathcal{G}^{\alpha\beta}_a(1,2)\varSigma^{\beta\gamma}_a(2,3) = (\theta_1-\theta_3) \delta(\tau_1 - \tau_3) \delta^{\alpha\gamma}.
\eea
The first one depends on the particular hypergraph while the second one is independent of it. 

For simplicity, we again consider regular hypergraphs. In this case, it is straightforward to show that there exists a uniform replica-diagonal solution, $\varSigma_a^{\alpha\beta}=\varSigma \delta^{\alpha\beta }$ and $\mathcal{G}_a^{\alpha\beta} =\mathcal{G}\delta^{\alpha \beta }$, satisfying,
 \bea 
 \varSigma(1,2) = kq \braket{C^2} \mathcal{G}^{q-1}(1,2) & = J  \mathcal{G}^{q-1}(1,2)\\
 D_{\theta_3} \mathcal{G}(1,3) +\int d \tau_2 d \theta_2 \mathcal{G}(1,2)\varSigma(2,3) &= (\theta_1-\theta_3) \delta(\tau_1 - \tau_3) .
 \label{eq:susy_sp_eom}
 \eea 
With the choice of the normalization $\braket{C^2}=J/kq$, the equation of motion in the uniform sector reduces to the all-to-all case. Both $\mathcal{G}$ and $\varSigma$ depends on two Grassman variables. We can expand them as,
 \bea
 \mathcal{G}(1,2) = G^{\chi\chi}(\tau_1, \tau_2) - \theta_2 G^{\chi\phi}(\tau_1, \tau_2) +   \theta_1 G^{\phi \chi }(\tau_1, \tau_2) + \theta_1 \theta_2 G^{\phi \phi} (\tau_1, \tau_2 ) \\
 \varSigma(1,2) = \theta_1\theta_2\Sigma^{\chi\chi}(\tau_1, \tau_2) + \theta_1 \Sigma^{\chi\phi}(\tau_1, \tau_2) +   \theta_2 \Sigma^{\phi \chi }(\tau_1, \tau_2) + \Sigma^{\phi \phi} (\tau_1, \tau_2 )
 \eea
From Eq. \ref{eq:susy_sp_eom}, one can derive the equation of motion for each component. 

The Green's functions with superscript $\chi \phi$ or $\phi \chi$ vanish for states with definite fermion parity or for a statistical mixture of states with definite fermion parity. In this case, the equation of motion is much simplified, 
 \bea
 -1 - G^{\phi \phi, -1}(\omega) &= \Sigma^{\phi\phi} (\omega), \ \ \  \Sigma^{\phi \phi} (\tau) = J G^{\chi\chi, q-1} (\tau)  \\
 -i\omega - G^{\chi \chi, -1}(\omega) &=\Sigma^{\chi\chi}(\omega), \ \ \ \Sigma^{\chi\chi}(\tau)=(q-1)J G^{\chi\chi,q-2}(\tau)  G^{\phi\phi}(\tau).
 \eea   

In the low energy limit, the solution is~\cite{Fu2016}
 \bea
 G^{\chi\chi}(\tau_{12}) = \frac{b\,\text{sgn}(\tau_{12})}{\left | \tau_{12} \right|^{2\Delta }}, \quad G^{\phi\phi} (\tau_{12}) = \frac{b\,\text{sgn}(\tau_{12})}{q\left | \tau_{12} \right|^{2\Delta +1  }},
 \eea
where $2\Delta =1/q$ and $J b^q = \tan \pi \Delta/(2\pi ) $. The solution can also be written in a more compact form,
\bea
\mathcal{G} = \frac{b\,\text{sgn}(\tau_{12})}{\left | \braket{1,2} \right|^{2\Delta }},
\eea
where $\braket{1,2}$ stands for $\tau_{12} -\theta_1\theta_2$ following the notation in~\cite{Murugan2017}.
 
For general non-regular hypergraphs, there is no longer a uniform solution. However, at low temperature one can show that $\gamma_a^{-1}\mathcal{G}_a$ and $\gamma_a \varSigma_a$ are uniform and are the same as the supersymmetric conformal solutions presented in~\cite{Fu2016}. Recall that the $\gamma_a$s are the factors defined in Eq.~\ref{eq:gamma} that regularize the hypergraph.

\subsubsection{Averaging over prunings}

We follow the same approach in Sec. \ref{sctn:ssyk_pruning} to get the random graph averaged action,
\bea
\overline{I_E}=& -\frac{1}{2}\sum\limits_a \int d \tau  d \theta  \Phi D_{\theta }\Phi + \frac{1}{2} \sum\limits_a \int d \tau_1 d \tau_2 d \theta_1 d \theta_2 \varSigma_a(1,2) \left(\Phi_a(1)\Phi_a(2) -\mathcal{G}_a(1,2) \right ) \\
&- p\sum\limits_A \exp\left( \frac{J}{2kq}\sum\limits_{\alpha\beta} \int d \tau
   _1 d \tau_2 d \theta_1 d \theta_2\mathcal{G}^{\alpha\beta}_{a_1}(1,2) ...\mathcal{G}^{\alpha\beta}_{a_q}(1,2) \right).
\eea
Again, the pruning probability is $kq (q-1)!N^{1-q}$. Similar to the non-susy sparse model, the interacting part of the equation of motion is modified to,
\bea
\varSigma _a =\frac{(q-1)!}{N^{q-1}} J \sum\limits_{a_2...a_q\neq a}\exp\left( \frac{J}{2kq}\sum\limits_{\alpha\beta} \int d \tau
   _1 d \tau_2 d \theta_1 d \theta_2\mathcal{G}^{\alpha\beta}_{a}(1,2)\mathcal{G}^{\alpha\beta}_{a_2}(1,2) ...\mathcal{G}^{\alpha\beta}_{a_q}(1,2) \right) \mathcal{G}_{a_2}...\mathcal{G}_{a_q}
\eea
It admits a uniform replica-diagonal solution satisfying,
\bea
\varSigma_s(1,2) = J \exp \left( \frac{mJ}{kq} \int d\tau_1 d\tau_2  d\theta_1 d\theta_2  \mathcal{G}_s^q(1,2)\right)\mathcal{G}_s^q(1,2).
\eea
where $m$ is the replica number.  In the physical limit $m \rightarrow 0$,  the extra factor vanishes and the equation reduces to that of the fully connected SUSY SYK model. Our comments about fluctuations in Sec.~\ref{sctn:ssyk_pruning} are relevant here as well, and it would be interesting to carry out the analysis in future work.

\section{Simulation of the sparse models}
\label{sctn:smltn}
\subsection{Classical simulation of sparse SYK}

In this section, we discuss the simulation of sparse SYK on a classical computer using exact diagonalization. The Hamiltonian of an all-to-all SYK model is much denser than Hamiltonians in geometrically local systems. It contains about $e^{N/2}N^q/q!$ nonzero entries, which should be compared with $e^{N/2} N$ entries for typical local systems. This density drastically increases the cost of all standard numerical methods and thus strongly restricts the system sizes that those methods can access. As we demonstrated in Sec. \ref{sctn:ssyk} and \ref{sctn:susy_ssyk}, the sparse model shares similar physical properties with the all-to-all case, yet the number of terms in the Hamiltonian decreases from $\sim N^q$ to $\sim N$. The sparse model is still non-local, but the numerical cost of some methods, such as exact diagonalization, become the same as in geometrically local systems.

Another obstacle within the context of exact diagonalization is the extremely long times and large storage required to fully diagonalize the Hamiltonian and obtain the full spectrum. However, in practice, the full spectrum is not really needed if we are only interested in local observables in thermal equilibrium. A direct approach to the equilibrium state is also challenging, because the density matrix $\rho =e^{-\beta H}/Z$ is typically not sparse which again limits the number of fermions one can simulate. However, at large $N$, typical states at a given energy density recover the physics of the equilibrium density matrix, at least as far as local observables are concerned. This is very helpful since we only need to access a typical pure state (a vector) instead of the equilibrium density matrix (a matrix) to calculate observables in thermal equilibrium.

To be explicit, we have an approximate equality,
\bea
\braket{O}_\beta = \frac{1}{Z} \text{tr} (\rho^{\frac{1}{2}} O \rho^{\frac{1}{2}}) \sim \frac{1}{\braket{\psi_\beta | \psi_\beta}}\braket{\psi_\beta|O|\psi_\beta},
\eea
where $\ket{\psi}=\rho^{1/2}\ket{\psi_0}$ and $\ket{\psi_0}$ is a random state in the Hilbert space. The imaginary time evolution of a random initial state can be realized within the standard Krylov space method, which does not require an explicit construction of the Hamiltonian but merely the action of the Hamiltonian on pure states. This approach allows us to take full advantage of the sparseness of the Hamiltonian. It is also straightforward to parallelize the algorithm to further increase speed: the Hamiltonian is the sum of order $N$ terms, each of which is a simple simple Pauli string, so each term can be applied to a state independently using different cores before the summing contributions from different cores at the end. 

One particularly interesting observable is the average energy in thermal equilibrium. From the temperature dependence of the energy, we can also extract the thermal entropy by integrating the thermodynamic relation $TdS =d E$ down from the high temperature,
\bea
\frac{S{(\beta)}}{N} = \frac{\log 2}{2} + \frac{1}{N} \left ( E(\beta)\beta - \int_0^\beta  E d\beta \right ),
\label{eq:entropy}
\eea
where $\log 2/2$ is the entropy at infinite temperature. Combining a typical state with real time evolution, one can also access time-dependent correlation functions of various operators.

\begin{figure}
\includegraphics[height=0.45\columnwidth, width=0.45\columnwidth]
{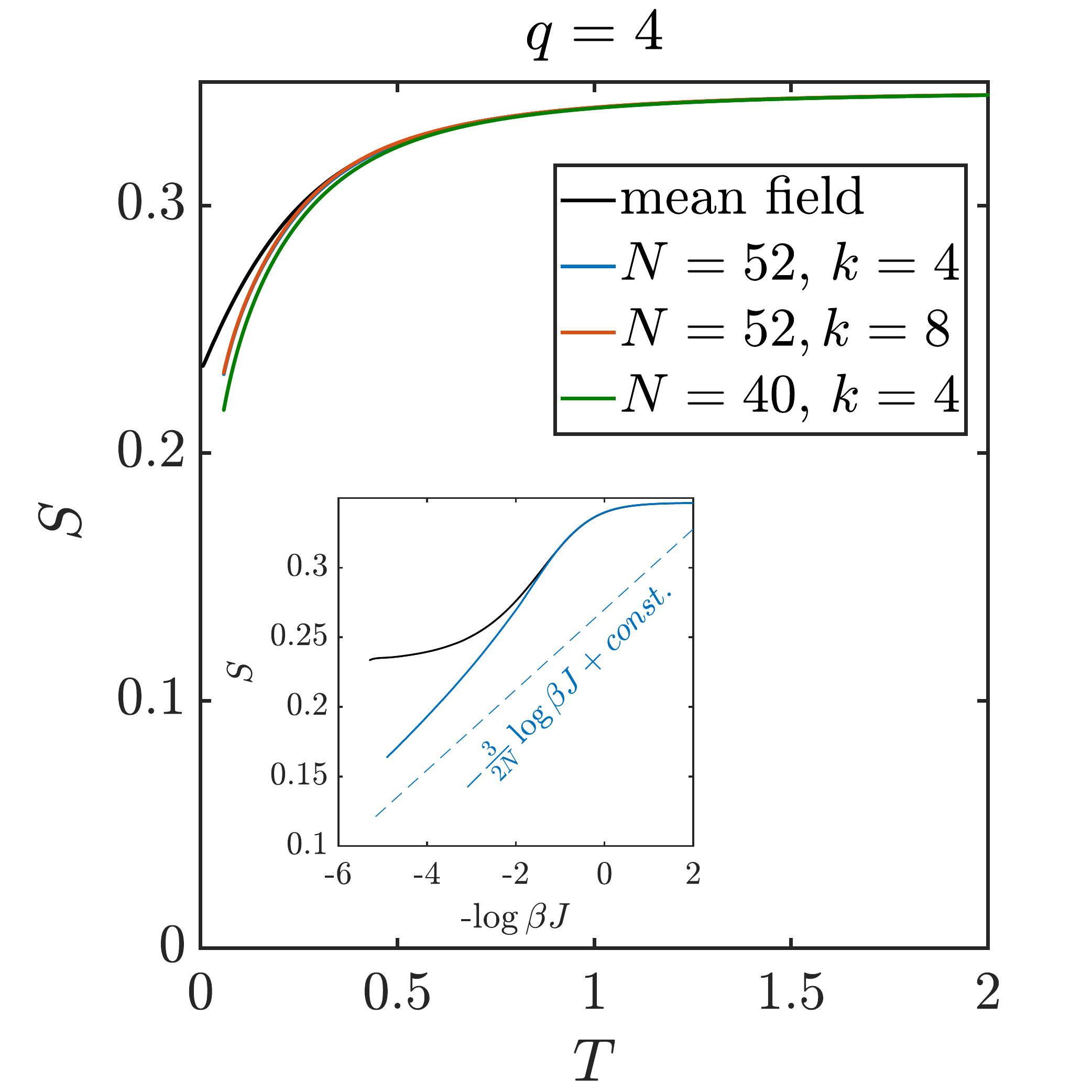}
\includegraphics[height=0.45\columnwidth, width=0.45\columnwidth]
{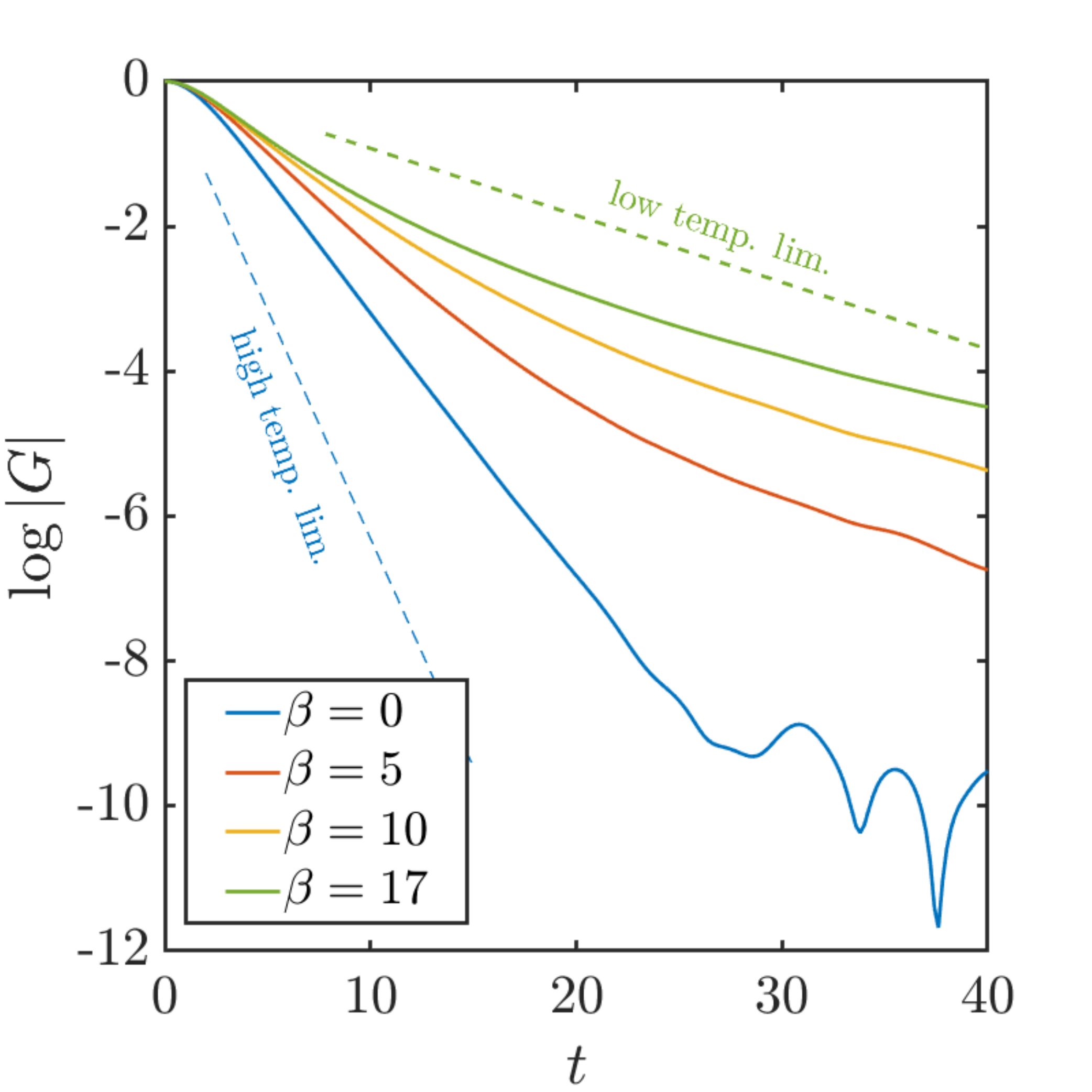}
\caption{(a) The entropy of the sparse $q=4$ SYK model as a function of temperature for different sparseness $k$ and majorana number $N$. The results are obtained using Eq. \ref{eq:entropy} from averaging 10 realizations of random coupling and random pruning. We observe the fluctuation across different realizations is intriguingly small  in the entropy curve. The deviation from the large-$N$ mean field solution is not from the finite-$k$ but a finite-$N$ effect. The lowest temperature is $\beta J=17$. The inset presents one specific realization for much lower temperature ($\beta J \sim 100$). It clearly demonstrates that the deviation is logarithmic in $\beta J$ with the coefficient $3/2N$, consistent with the prediction of the low-energy Schwarzian theory.   (b) The real-time green function from one particular realization at different temperatures. They exhibit a window of exponential decay. The decay rate agrees with both the low temperature analytical value $\frac{\pi }{2}T$ and the high temperature numerical value $0.6307 J$~\cite{Roberts2018} for the all-to-all connected SYK model. }
\label{fig:q4}
\end{figure}

Using these procedures, we are able to simulate the $q=4$ sparse SYK model to obtain local observables in thermal equilibrium for $N$ up to $52$. This value of $N$ is straightforward to achieve in the sparse model, and going to even larger $N$ is possible. Previous studies~\cite{Gur_Ari_2018,Kobrin2020} have reached up to $60$ fermions in the non-sparse model with enormous computer time~\cite{Kobrin2020}, and one could certainly go further with the sparse model. The results are presented in Fig. \ref{fig:q4}, where we show the temperature dependence of the entropy for various $N$ and $k$. The results are compared to the large $N$ all-to-all solution, obtained by numerically integrating the large $N$ equation of motion. To obtain this data, we generated hypergraphs from random pruning and averaged $10$ random realizations of the hypergraph structure and $J$ the couplings. Remarkably, the temperature dependence of the entropy at $N=52$ is almost identical for $k=4$ and $k=8$ down to the lowest temperatures studied. The $N=52$ data is also closer to the large $N$ reference case than the $N=40$ data. This suggests that the dominant source of deviation from the all-to-all large $N$ result is due to finite $N$ effects as opposed to $1/k$ effects. In particular, while there are certainly $1/k$ corrections to the fully connected result, it appears empirically that these corrections are quite small for all the energies we accessed, as expected from the analysis in Sec. \ref{sctn:ssyk} and \ref{sctn:susy_ssyk}.

For one particular realization with $N=52$ and $k=4$, we lowered the temperature down to $\beta J \sim 100$. The inset in Fig.~\ref{fig:q4} shows that the deviation of the data from the large $N$ fully connected solution follows the formula $-\frac{3}{2N} \log \beta J +\text{constant}$. This expression is the expected $1/N$ correction to the entropy in the fully connected model, so its appearance in the sparse data is further evidence that the low energy Schwarzian theory is present in the sparse model when $k=4$. We also calculate real time Green functions for one particular realization at a variety of temperatures. The Green functions exhibit a window of exponential decay, and the temperature dependent decay rate agrees with the analytical all-to-all value $\frac{2\pi}{\beta q}$ at low temperature~\cite{Maldacena2016} and the numerical all-to-all value at high temperature~\cite{Roberts2018}. This provides extra evidence that the sparse model recovers the interesting physics of its all-to-all counterpart. 

\begin{figure}
\includegraphics[height=0.45\columnwidth, width=0.45\columnwidth]
{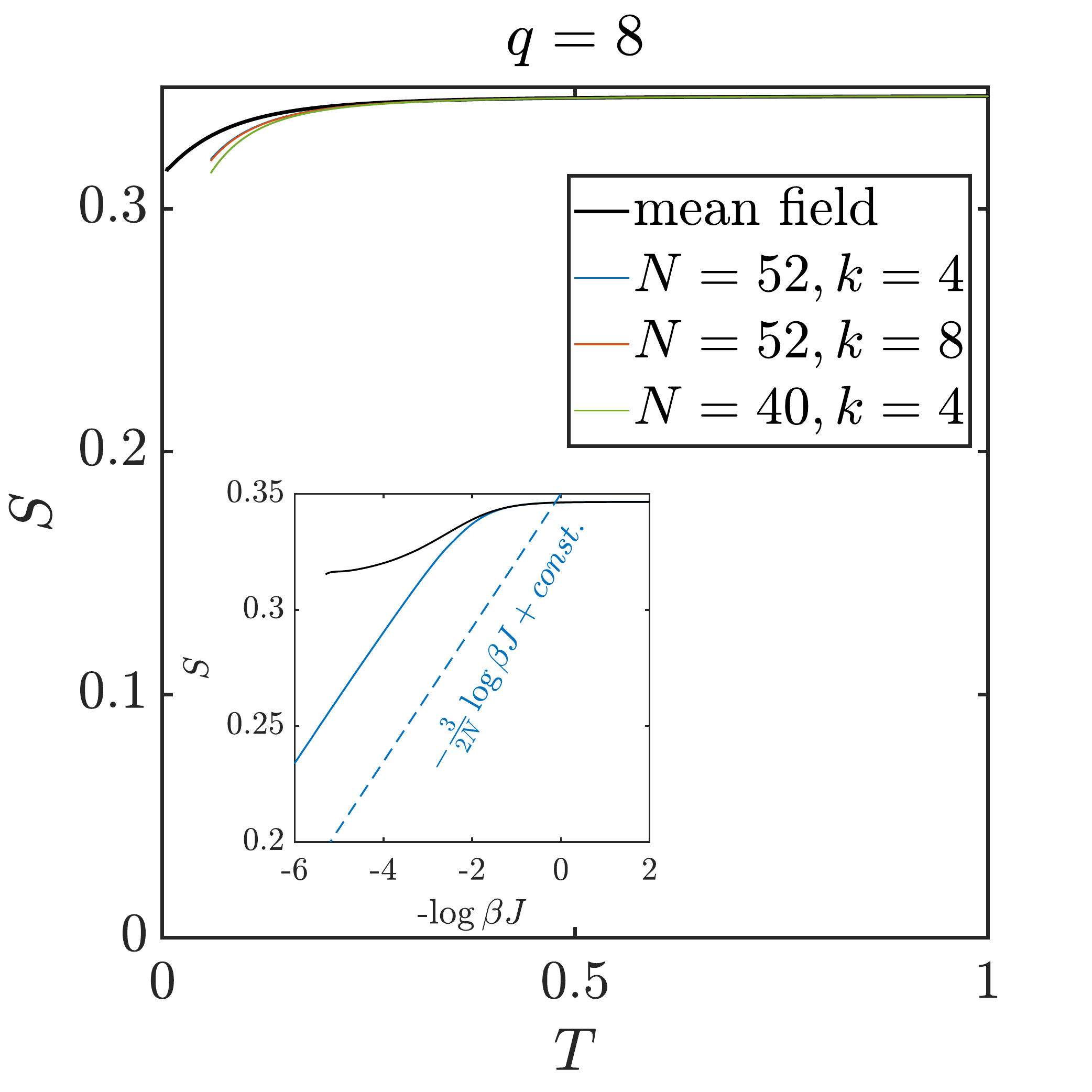}
\includegraphics[height=0.45\columnwidth, width=0.45\columnwidth]
{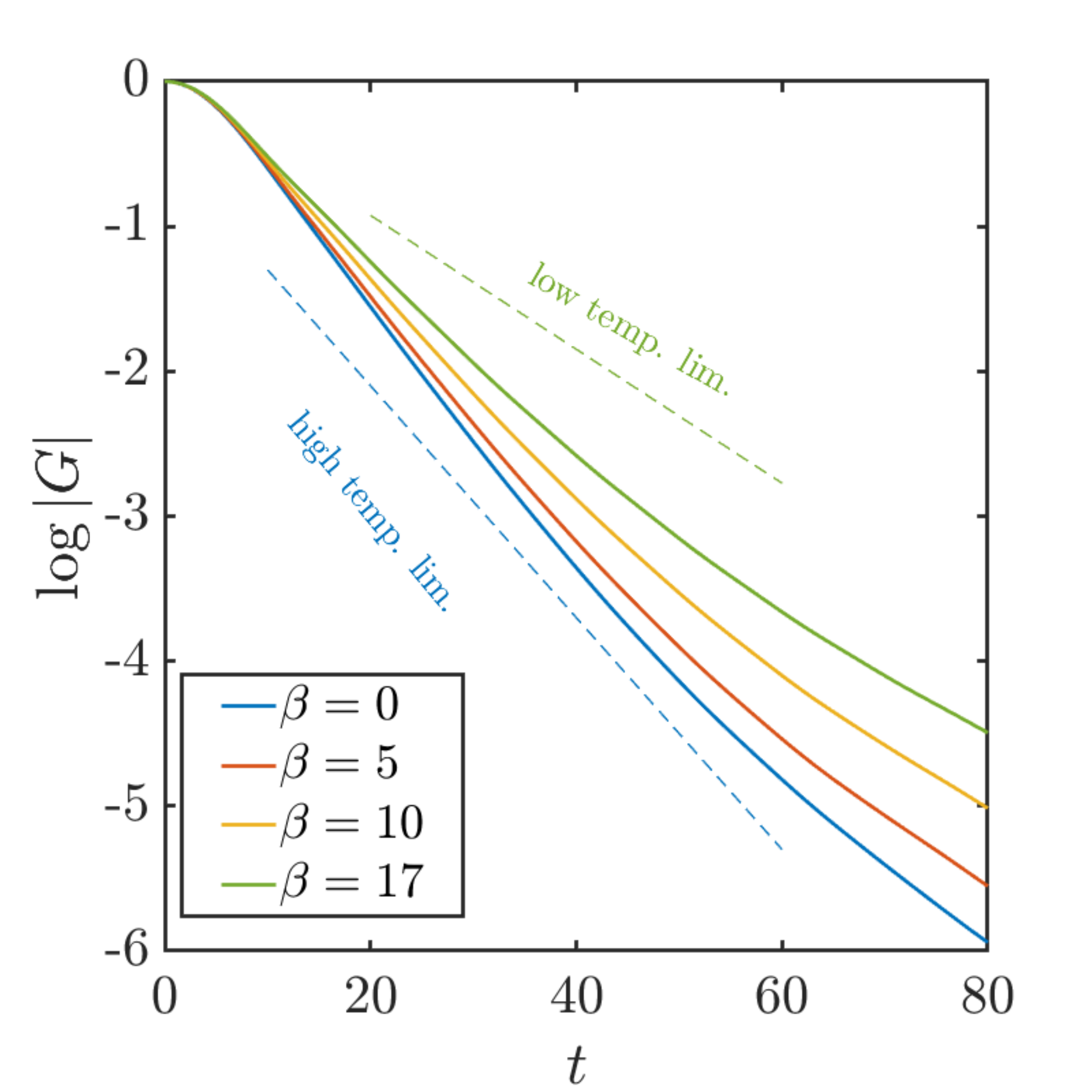}
\caption{(a)  The entropy of the sparse q=4 SYK model as a function of temperature for different sparseness $k$ and majorana number $N$. The results are obtained using Eq. \ref{eq:entropy} from averaging 10 realizations of random coupling and random pruning. We observe the fluctuation across different realizations is intriguingly small  in the entropy curve. The deviation from the large-$N$ mean field solution is not from the finite-$k$ but a finite-$N$ effect. The lowest temperature is $\beta J=17$. The inset presents the entropy curve for one specific realization down to  much lower temperature ($\beta J \sim 800$), demonstrating that the deviation from the large-$N$ case is consistent with the finite-$N$ effect obtained from the low-energy Schwarzian theory. (b) The real-time green function from one particular realization at different temperatures.  }
\label{fig:q8}
\end{figure}

We can also consider larger values of $q$. The Hamiltonian of the all-to-all SYK model becomes much more dense as $q$ increases. For example, SYK model with $q=8$ has $\sim N^8$ terms. Therefore, most large-scale numerical simulations focus on the $q=4$ case~\cite{Kobrin2020}. On the other hand, one of the advantage of the sparse version is that simulation complexity does not increase with $q$, providing a route to access the physics at larger $q$. We simulate the sparse model for $q=8$ up to $N=52$, with results presented in Fig.~\ref{fig:q8}. Similar to $q=4$, the value of $k$ hardly affect the temperature dependence of the entropy, and the deviation from the large-$N$ solution agrees with that predicted from the Schwarzian theory. For completeness, we also study the supersymmetric variation of the sparse SYK model for both $\mathcal{N}_{sys}=1$ and $\mathcal{N}_{sys}=2$ (which we did not explicitly define, but which follows the same structure as the $\mathcal{N}=1$ case), at $N=52$ and $N=26$, respectively. In the latter case, $N$ counts the number of the complex fermions, so the Hilbert space dimension is the same as that of $52$ Majoranas. The entropy curve of the $\mathcal{N}_{sys}=2$ case is above the analytic results because its ground state degeneracy is protected for any $N$ as shown by the Witten index~\cite{Fu2016}. 

\begin{figure}
\includegraphics[height=0.45\columnwidth, width=0.45\columnwidth]
{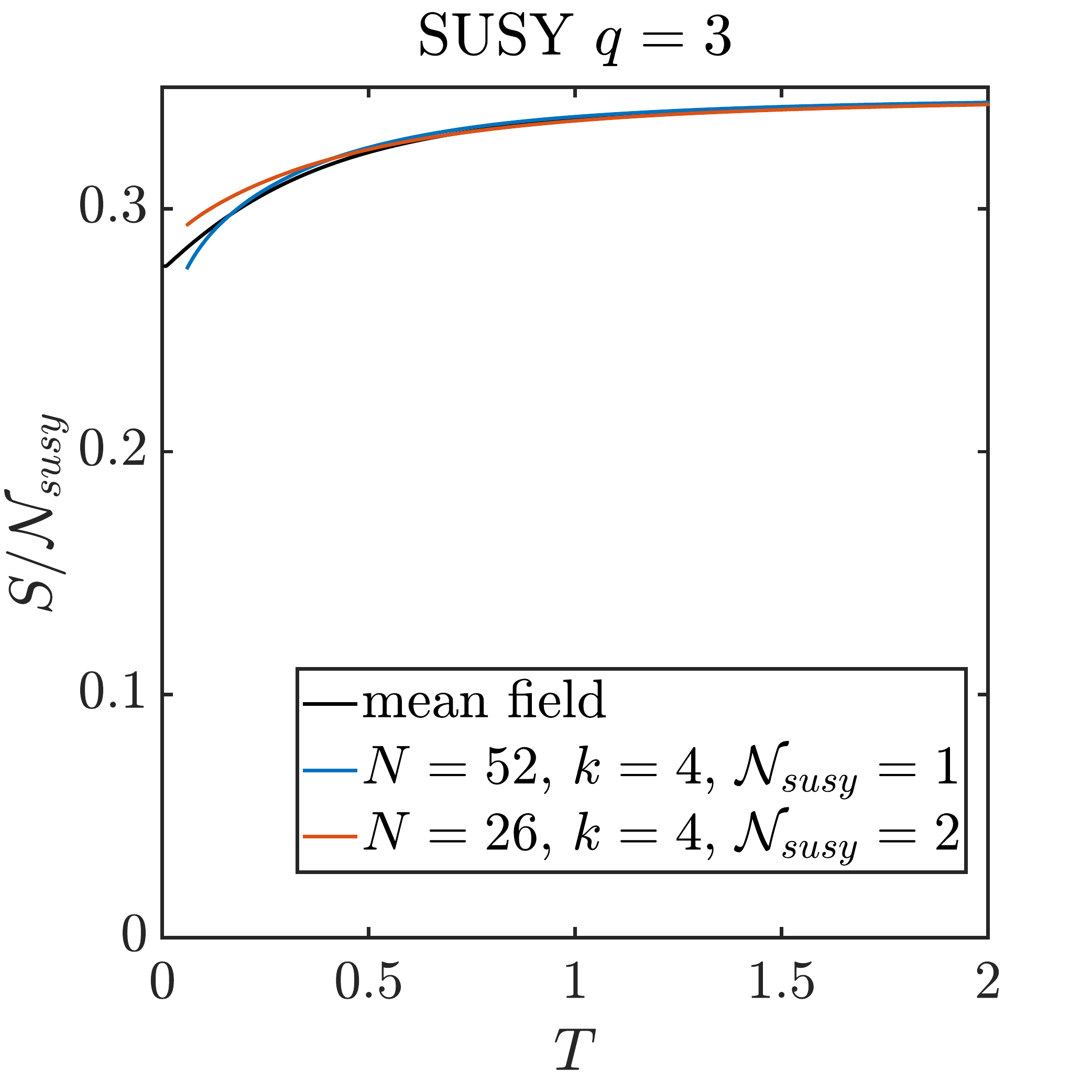}
\caption{The temperature dependence of the entropy for the SUSY $\mathcal{N}_{susy}=1$ and $\mathcal{N}_{susy}=2$ sparse  SYK model at $q=3$. }
\label{fig:susy}
\end{figure}

\subsection{Quantum simulation of sparse SYK}

In this section, we discuss the complexity of simulating the sparse SYK model on a quantum computer.
We consider two common quantum algorithms for Hamiltonian simulation: the simulation algorithm based on product formulas~\cite{Lloyd1996} and the qubitization algorithm~\cite{Low2016}. We discuss both schemes in the context of simulating the sparse SYK model and show that the cost is drastically reduced relative to the all-to-all SYK model.

\subsubsection{Product formulas}

Given a Hamiltonian $H=\sum_{i=1}^L H_i$ containing $L$ terms, digital quantum simulation can be achieved by using product formulas to approximate the time evolution operator $e^{-  i H t }$. 
The well-known Lie-Trotter formula $\mathcal{F}_1(dt) =e^{-i H_1 dt}e^{-i H_2 dt}\cdots $ is a first-order product formula with error
\bea
e^{-i H dt } -\mathcal{F}_1(dt) = \frac{1}{2}\sum \limits _{i<j }[H_i, H_j] dt^2 +O(dt^3).
\eea
From it, the time evolution operator can be approximated by $\mathcal{F}_1^n(t/n)$, with error bounded by
\bea
||e^{-i H t} -\mathcal{F}_1^n(t/n)|| =\mathcal{O} \left(\sum_{i,j}||[H_i, H_j]|| \frac{t^2}{n}\right).
\eea
Note that the accuracy increases as $n$ grows. The most general $p$th-order product formula takes the form
\bea
\mathcal{F}_p(dt)= \prod \limits_{i=1}^{\Gamma(p)}  \left ( \prod\limits_{j=1}^{L} \exp\left (-i a_{i,j} H_{\pi_i (j) }dt \right ) \right )=e^{-iHdt}+O(dt^{p+1}),
\label{eq:product_formula}
\eea
where $\pi_i$ denotes a permutation and $\Gamma(p)$ represents the number of stages, which usually increases with the order $p$. Recent work~\cite{Childs2020} shows that, analogous to the first-order formula, the error of the $p$th-order formula is given by 
\bea
||e^{-i H t} -\mathcal{F}_p^n(t/n)|| =\mathcal{O} \left( \alpha_p \frac{t^{1+p}}{n^p}\right),\quad
\alpha_p =\sum_{i_1\cdots i_{p+1}}||W_{i_1\cdots i_{p+1}}||,
\eea 
where $W$ is the nested commutator defined recursively through $W_{i_1\cdots i_{p+1}}=[H_{i_{p+1}}, W_{i_1\cdots i_p}]$. Therefore, the number of Trotter steps $n$ required to achieve precision $\epsilon$ is 
\bea
n=\mathcal{O}\left(\alpha_p^{\delta}t^{1+\delta }\epsilon^{-\delta}\right),  \ \ \ \delta=\frac{1}{p},
\eea
and the total complexity is the gate complexity to realize each $\mathcal{F}_p(t/n)$ multiplied by $n$. 

Now we consider digital quantum simulation of the sparse SYK model. The key step to obtain the complexity is to estimate $\alpha_p$ from the nested commutators.  In the SYK model, every term in the Hamiltonian is a majorana string. As a result, the nested commutator $W_{i_1\cdots i_{p+1}}$ is a majorana string with at most $(p+1)q-2p$ majorana operators, which is noncommutative with $\sim ((p+1)q-2p ) kq$ terms in the Hamiltonian that contributes to $\alpha_{p+1}$. Based on this, we estimate that in the sparse SYK model, 
\bea
\alpha_p =  \mathcal{O}\left(N J^{1+p}  2^{-((p+1)q-2p)/2} k^{\frac{p+1}{2}} q^{\frac{p-1}{2}} \prod \limits_{i=1}^p (i q-2i +2)\right) = \mathcal{O}\left(N J^{1+p}  2^{-((p+1)q-2p)/2} k^{\frac{p+1}{2}} q^{\frac{3p-1}{2}}\right). \\
\eea
As a result, 
\bea
n =\mathcal{O}\left(  N^{\delta }2^{-q/2 (1+\delta) }k^{(1+\delta)/2 }q^{(3-\delta )/2}\epsilon^{-\delta}(Jt)^{1+\delta})\right) =\mathcal{O}(N^\delta k^{(1+\delta)/2}q^{1-\delta} \epsilon^{-\delta}(\mathcal{J}t)^{1+\delta}).
\eea
It is conventional to use $\mathcal{J}=\sqrt{q2^{(1-q)}}J$ instead of $J$ in the large $q$ limit. 

Next we estimate the gate complexity to realize the unitary operator  $\mathcal{F}_p(t/n)$. According to the definition, a $p$th-order product formula for the sparse SYK model is a product of  $\mathcal{O}(kN)$ unitary operators $e^{-i a H_b t }$, where $H_b$ takes the form of a majorana string $J_{a_1\cdots a_q} \chi_{a_1}\cdots \chi_{a_q}$. We map the majorana operators to the Pauli operators using the Bravyi-Kitaev transformation \cite{Bravyi2002} so that a single majorana operator maps to a product of $\mathcal{O}(\log N)$ Pauli operator. This is much more efficient than the conventional Jordan-Wigner transformation where a single majorana operator maps to a product of $\mathcal{O}(N)$ Pauli operators. After the transformation, $H_b$ becomes a Pauli string with $\mathcal{O}(q\log N)$ Pauli operators. The gate complexity to implement the unitary $\exp(-i a H_b dt)$ is $\mathcal{O} (q\log N)$, and the gate complexity to implement $\mathcal{F}_p(t/n)$ is $\mathcal{O}(kq N \log N) $.

Adding all these together, we obtain  the total gate complexity to simulate the Hamiltonian time evolution $e^{-i H t}$ with precision $\epsilon$
\bea
\mathcal{C} =n \mathcal{C}(F_p (t/n) ) = \mathcal{O} (\epsilon^{-\delta }
k^{(3+\delta)/2}q^{2-\delta}  N^{1+\delta}\log N  (\mathcal{J}t)^{1+\delta} ).
\eea
When $k$ and $q$ is order 1, the gate complexity is $\mathcal{O} (\epsilon^{-\delta} N^{1+\delta} \log N (\mathcal{J}t)^{1+\delta})$. 
In comparison, simulating the all-to-all model with product formulas would have complexity
\begin{equation}
    \mathcal{O}\left(e^{(3+\delta)q/2}2^{-(1+\delta)q/2} N^{2q-1+(q+1)\delta} q^{(\frac{q}{2}+\frac{3}{4})(-1+\delta)} (Jt)^{1+\delta}\epsilon^{-\delta}
    \log N\right);
\end{equation}
see Appendix~\ref{sctn:all2all} for details. We have thus seen that the sparse SYK model is much easier to simulate on a quantum computer than its all-to-all counterpart.

Note that although we have focused on using product formulas to simulate the sparse SYK model, similar result may also be achieved using the algorithm based on the Lieb-Robinson bounds \cite{HHKL18}. The idea of simulating a randomly sparsified Hamiltonian is also advocated in \cite{Ouyang2020compilation}, although their bound is not tight for the sparse SYK model defined from regular hypergraphs. Finally, it is possible to use alternative femionic encodings such as \cite{Jiang2020optimalfermionto} to implement the sequence of elementary exponentials in product formulas.

\subsubsection{Qubitization}
Now we consider a different quantum algorithm named qubitization~\cite{Low2016} to simulate the Hamiltonian dynamics. The idea is to encode the Hamiltonian in the physical space by a unitary operator  $U$ acting on an enlarged Hilbert space, a tensor product of the physical Hilbert space and an ancilla Hilbert space. The input is the unitary operator $U$ and two states $\ket{G_L}$ and $\ket{G_R}$ in the ancilla space. They are defined as follows
\bea
&U= \sum\limits_l  \ket{A}\bra{A}_{\text{anc}} 2^{q/2} \chi_{a_1}\cdots \chi_{a_q}\ \ 
&\ket{G_L} =\frac{1}{\sqrt{\sum J_A^2}} J_A \ket{A}_{\text{anc}}, \ \ \ \ket{G_R}= \frac{1}{\sqrt{L}} \sum \ket{A}_{\text{anc}} .
\eea
One can easily verify that
\bea
\bra{G_L} U \ket{G_R} = \frac{1}{\lambda} H,  \ \ \lambda = \frac{L}{2^{q/2}} \sqrt{\frac{1}{L} \sum J_A^2} = \sqrt{\frac{k}{q} }\frac{N J}{2^{q/2}}.
\eea
The advantage of using different $\ket{G_L}$ and $\ket{G_R}$, a scheme called asymmetric qubitization~\cite{Babbush2019}, is that they are relatively simpler to make than the state used in the symmetric version. Using $U$, $\ket{G_L}$ and $\ket{G_R}$ $n$ times, the qubitization algorithm maps the Hamiltonian simulation problem to optimal control of a sequence of single-qubit rotations \cite{Low2017}. The details can be found in \cite{Low2016, Low2017, Babbush2019}. The output of the algorithm is an operator $F_n(H)$ only acting on the physical space, which is an order $n$ polynomial of $H$. Via optimal control, the polynomial is made equal to the the truncated Jacobi-Anger series, 
\bea
F_n (H/\lambda) = J_0(-\lambda t) +2 \sum \limits_{m=1}^q i^n J_n (-\lambda t) T_m (H/\lambda)\approx \exp(-i H t),
\eea
where $T_m$ is the Chebyshev polynomials. 
As $n$ increases, the series rapidly converges to $e^{-i Ht}$ with deviation $ ||F_n(H/\lambda) - \exp(-i H t)|| < \mathcal{O}((\frac{e \lambda t}{2n})^n)$. Therefore the required number of queries  $n$ needed to achieve accuracy $\epsilon$ is $\mathcal{O}(\lambda t + \log \epsilon^{-1} )$.

Let $\mathcal{C}$ be the gate complexity to implement the unitary $U$ and the states $\ket{G_{L/R}}$.  The dominant cost is from $U$ which can be implemented by sawtooth circuits~\cite{CMNRS18,Babbush2018} and requires $\mathcal{O} (kqN\log N)$ gates. It is worth noting that because of the sparseness of the sparse SYK Hamiltonian we consider, $U$ here does not directly factorize as it does for the full SYK model~\cite{Babbush2019}. As a result, we iterate over all $kN$ terms from the Hamiltonian and there is an additional $\log N$ factor associated with the length of the Pauli string of each interaction term after the Bravyi-Kitaev transformation. This may be improved by using an implementation of $U$ similar to that of \cite{Babbush2019} and a more complicated procedure to prepare the state $\ket{G_L}$. We leave a detailed study of such an improvement as a subject for future work. The leading contribution of $N$ and $t$ to the total complexity of the algorithm is 
\bea
\mathcal{C} = \mathcal{O} (kq N\log N \lambda t) =\mathcal{O} (k^{3/2}N^2\log N \mathcal{J} t).
\eea
This is a significant reduction of the cost $\mathcal{O}\left(\left (N^{q+3}/q! \right )^{1/2} \mathcal{J} t \right)$ for simulating the full SYK model using qubitization.

\section{Discussion}
\label{sctn:discussion}

Let us summarize the results contained in this paper. First, we defined sparse versions of the SYK model where the number of terms is equal to $kN$, either exactly (using regular hypergraphs) or on average (using randomly pruned hypergraphs). We also defined analogous sparse versions of supersymmetric SYK models. With a proper normalization of the coupling variance, it is straightforward to establish that these models enjoy a $1/k$ expansion and that they recover the physics of the fully connected model as $k\rightarrow \infty$.

Second, we developed a path integral approach to analyze these sparse models. Focusing on the case of regular hypergraphs, we showed that the saddle point equations admit a uniform solution that is replica symmetric and diagonal and that the saddle point action is the same as in the fully connected model. However, unlike the fully connected model, fluctuations around the saddle point are not $1/N$ suppressed and do modify the free energy at leading order in $N$. To quadratic order in fluctuations and in the low temperature limit, we showed that there was a single soft mode, the same mode as in the fully connected model, and that all other modes were gapped.

Third, we showed that sparsity dramatically simplifies both numerical exact diagonalization calculations and quantum simulations. We presented numerical data for up to $N=52$ fermions that were consistent with our path integral arguments. We also computed the gate complexity of quantum simulations using two different algorithms, product formulas and qubitization.

Based on these results, we conclude that the sparse model exhibits a maximally chaotic gravitational sector at low energy which is governed by the same effective action as in the fully connected model, up to a possible renormalization of parameters, provide $k > k_{\min}$. Although we do not precisely determine $k_{\min}$, the numerical data suggest that $k_{\min} < 4$. The physical picture is moreover clear: the gravitational sector arises as a consequence of spontaneous and explicit breaking of emergent conformal symmetry and this pattern of symmetry breaking, which is a global phenomenon, persists in the sparse limit. We further conclude that, for the purposes of studying holographic models of quantum gravity using classical or quantum simulations, the sparse model is superior to the fully connected model in terms of resource costs.

Nevertheless, the sparse model is not solvable to the same degree as fully connected SYK. In particular, there are certainly $1/k$ corrections to various physics quantities that are interesting to compute in future work. We computed the first order term at high temperature and found a correction of order $1\%$ for $k=4$ and $q=4$. Furthermore, empirically, the effects of varying $k$ are quite small even for $k$ as small as $4$. It would be very interesting to understand these corrections better. Perhaps a renormalization group analysis of some sort would be helpful. In particular, it seems conceivable that the sparsity is in some sense an irrelevant perturbation at low energy, e.g. that $k$ effectively flows to infinity.

The sparse model also fails to self-average (with respect to the disorder) to the same degree as the fully connected model. We discuss this extensively in Appendix~\ref{sctn:flctn} and simply summarize our conclusions here. The partition function of the sparse model is not self-averaging even at high temperature. This is a simple consequence of the fact that the Hamiltonian contains only order $N$ terms, rather than order $N^q$ terms. However, from a variety of analytical arguments and numerical data, we find that the energy density is self-averaging at large $N$. Moreover, as with the $1/k$ corrections, the practical size of disorder fluctuations is quite small at large $N$.

It is also interesting to note that the $q=2$ case is different from $q\geq 4$. In particular, the $q=2$ model does not have a gravitational sector. However, analytics and large scale numerics are easier since the physics is determined by the single particle energy levels. These levels are obtained from the spectrum of a sparse anti-symmetric matrix with Gaussian random entries. A preliminary numerical investigation suggests that there is a strong enhancement of the density of states near zero energy and a strong deviation from the semi-circle law, especially for low $k$. Hence, the sparse $q=2$ model may be significantly different from its non-sparse version.

\subsection{What is a legal many-body Hamiltonian?}

We now turn to some comments on our main results. Besides ``practical'' questions concerning the minimal resources needed to simulate quantum gravity, the sparse models also suggest to broaden our notion of what is a legal or acceptable many-body system. We propose that any family of Hamiltonians indexed by system size $N$ should be regarded as legal if there is a family of efficiently specified quantum algorithms indexed by $N$ with gate complexity polynomial in $N$.

Certainly any local lattice model falls into this class, including, for example, all standard models in condensed matter physics and lattice regularizations of gauge theories and other field theories. Fully connected models also fall into this class provided that the interactions are few-body, meaning the analog of the $q$ parameter is fixed as $N$ increases. The gate complexity may be higher than local models, but it is still polynomial in $N$. However, a model where $q$ grows with $N$, for example, the so-called double scaling limit of SYK~\cite{Berkooz2019} where $q\sim \sqrt{N}$, does not fall into this class. This is because the Hamiltonian contains a number of terms scaling  superpolynomially with $N$. Remarkably, the sparse versions of the double-scaled models do satisfy our definition of a legal many-body system, and since, with such a large $q$, the sparse physics should be very close to the fully connected limit, we have non-trivially increased the set of many-body systems that can be effectively quantum simulated.

Of course, our notion of a legal many-body system is somewhat arbitrary. Certainly digging rocks up will only provide so many different kinds of systems, and presumably nothing like a double-scaled SYK model. However, with the power of quantum simulation, we can effectively perform experiments on a much larger family of models. Our proposal is that a system is legal if it meets the sensible requirement that we can ``efficiently'' (with polynomial effort) simulate experiments on it.

\subsection{Extension to non-random models}

Beyond the SYK model, there are non-random models that are known to exhibit holography. Of course, the original examples of holography come from large $N$ gauge theories. More recently, the Gurau-Witten tensor models~\cite{Gurau2017,Witten2019} and their descendants have been shown to exhibit similar physics to SYK at large $N$ despite having no randomness. It is interesting to define sparse versions of these models. 

For the Gurau-Witten tensor model, there is a natural possibility. The conventional model is defined on $q=D+1$ real fermion fields $\psi_0, \cdots, \psi_D$~\cite{Witten2019}. Each fermion has $n^D$ components for an integer $n$. The theory possesses a large symmetry group, $G = O(n)^{D(D+1)/2}$, up to a discrete quotient. Each $\psi_a$ transforms as a real irreducible representation of $G$ according to the following pattern. For each unordered pair $a,b$ of distinct elements drawn from $\{0,\cdots,D\}$, introduce a group $G_{ab} = O(n)$. Up to a possible discrete quotient, the full symmetry group is then 
\begin{equation}
    G= \prod_{a<b} G_{ab} = O(n)^{D(D+1)/2}.
\end{equation}
The discrete quotient arises because a discrete subgroup of $G$ leaves all fermions invariant, so only the quotient by this discrete subgroup acts faithfully. The fermion $\psi_a$ transforms as a vector under $G_{ab}$ for all $b$ and transforms trivially under $G_{bc}$ when $b$ and $c$ are both different from $a$.

The Hamiltonian of the system is denoted by the symbol $i^{q/2} \psi_0 \cdots \psi_D$, which indicates the unique $G$ invariant contraction of all the fermion indices~\cite{Witten2019}. In detail, for each factor $G_{ab}$, exactly two fermions, namely $\psi_a$ and $\psi_b$, transform as vectors under $G_{ab}$. The relevant indices on $\psi_a$ and $\psi_b$ are therefore contracted in $\psi_0 \cdots \psi_D$. Remarkably, this model reproduces the leading large $N$ diagram expansion of SYK, although the subleading corrections are different.

The model itself appears rather rigid in its structure, and the only way we have found to sparsify the model does not preserve the full symmetry group. The construction proceeds as follows. Suppose $n= m n_0$ for two smaller integers $m$ and $n_0$ and consider a subgroup $O(n_0)^m \subset O(n)$. Setting $G'_{ab} = O(n_0)^m$, each fermion $\psi_a$ now forms a reducible representation of the $G'_{ab}$ factors. An example should clarify the structure. Consider fermion $\psi_0$, which has components $\psi_{0,i_{01} \cdots i_{0D}}$, where $i_{0b}$ denotes a vector index of $G_{0b}$. Regarded as a function of just $i_{01}$, say, the vector $v_{i_{01}} =  \psi_{0,i_{01} \cdots i_{0D}}$ is a direct sum of irreducible representations of $G'_{01} = O(n_0)^m$. Each irreducible representation in this sum transforms as vector under one $O(n_0)$ factor and is invariant under the rest. Hence, it is convenient to replace the index $i_{0b}$ running from $1$ to $n$ with a new index pair $i'_{0b} \ell_{0b}$ where $i'_{0b} = 1,\cdots,n_0$ denotes a vector index of $O(n_0)$ and  $\ell_{0b} = 1,\cdots,m$ indicates the $O(n_0)$ factor in $G'_{0b}$.

The original Hamiltonian $i^{q/2} \psi_0 \cdots \psi_D$ can now be viewed as a sum of invariants of the group $G' = \prod_{a<b} G'_{ab}$. The sum is labelled by the set $\{\ell_{ab}\}$ which indicates which $O(n_0)$ factors are non-trivially involved in an invariant. Based on this fact, an opportunity for sparsification presents itself. One can define a new Hamiltonian by deleting any number of the $G'$-invariant terms from the original Hamiltonian. The resulting Hamiltonian is no longer $G$-invariant, but it remains $G'$-invariant. Moreover, if one deletes enough terms, then the resulting Hamiltonian can be taken to the sparse limit. There are $m^{D(D+1)/2}$ possible labels $\{\ell_{ab}\}$ and $q m^D$ fermions, each with $n_0^D$ components. To keep $k q m^D$ terms, we must retain a fraction
\begin{equation}
    p = \frac{k q m^D}{m^{D(D+1)/2}} = \frac{k}{m^{D(D-1)/2}}
\end{equation}
of the terms. Hence, one can define a sparse model in the thermodynamic limit by taking $m$ to infinity with $k$ and $n_0$ fixed. The resulting Hamiltonian does have a large symmetry group, namely $G'$, but it is no longer invariant under the original $G$ symmetry. At this point, however, we know very little about this model. At large $m$, one possibility is that the physics of the full model is partially recovered and the symmetry is enhanced back to $G$. It would be very interesting to understand whether this occurs. For that purpose, quantum simulations of the model might be very useful given the paucity of analytical tools.

\textit{Acknowledgements}---We thank Edward Witten for helpful discussions and collaboration at an earlier stage of this work. BGS acknowledges support from the U.S. Department of Energy
(DOE), Office of Science, Office of Advanced Scientific Computing Research (ASCR) Quantum Computing Application Teams program, for support under fieldwork proposal number ERKJ347 and from the Simons Foundation via the It From Qubit collaboration. SX acknowledges the the support from University of Maryland supercomputing resources and Texas A\&M High Performance Research Computing. YS thanks Ryan Babbush for helpful discussions on quantum simulation. He is supported by the Google Ph.D. Fellowship program, ARO MURI, NSF (Grant No. CCF-1813814), the U.S. Department of Energy, Office of Science, Office of Advanced Scientific Computing Research, Quantum Algorithm Teams, Quantum Testbed Pathfinder (Award No. DE-SC0019040), Accelerated Research in Quantum Computing (Award No. DE-SC0020312) programs, the National Science Foundation RAISE-TAQS 1839204, and Amazon Web Services (name of program - AWS Quantum). The Institute for Quantum Information and Matter is an NSF Physics Frontiers Center. LS acknowledges support from the Simons Foundation via the It From Qubit collaboration.

Note added: This work has been in progress for more than two years, with most of the results discussed here presented in final form at a conference talk by one of the authors in September, 2019~\cite{Swingle2019}. Just prior to the appearance of our work, the paper~\cite{Garcia-Garcia2020} appeared which presents a study of the spectral statistics of energy eigenvalues in the sparse model based on the definitions and results presented in~\cite{Swingle2019}.

\bibliography{SYK,graph,quantum_computation,scrambling}

\appendix
\begin{appendices}

\section{Definitions and Properties of Graphs}
\label{sctn:graphs}

\begin{enumerate}
\item A graph is a collection of $N$  vertices and some number of edges connecting the vertices. Vertices and edges are sometimes called sites and links. The graphs we will consider have no edges connecting a vertex to itself, and at most one edge connects any two vertices. 

\item The degree of a vertex is the number of edges that emerge from that vertex. Graphs in which every vertex has the same degree $d$ are called $d$-regular. To be a little more precise, a family of graphs is $d$-regular if all its members are $d$-regular. In particular this means that $d$ is independent of $N$. 

\item A complete graph is one in which every pair of vertices is connected by an edge. Points, line-intervals, triangles, tetrahedrons are all complete graphs. The degree of every vertex in a complete graph has the maximum possible value, namely $d=(N-1).$ Complete graphs are the most dense graphs. By comparison $d$ regular graphs become  very sparse as $N$ grows.

\item Strict regularity is a very strong requirement which we will relax to allow a more general concept: statistical regularity which allows   fluctuations of bounded variation about the average  degree. The average degree will be  called $\bar{d}$. A statistically regular family is defined so that  both the average degree and the variance are independent of $N$. 

\item A graph is homogeneous if it looks the same from every point. Again we will allow a looser concept of statistical or approximate  homogeneity.

\item  We will use the term \it expander  graph \rm to mean that the graph is statistically both $d$-regular and homogeneous, and  has \it strong expansion. \rm

\item To define expansion, we arbitrarily select a vertex and work outward from it. Any other vertex has a graph distance from the selected vertex which is defined as the minimum number of edges connecting the two. An expander graph has the property that the number of vertices at a distance $L$ from the selected vertex grows exponentially with $L$. 

Since $N$ is finite a given graph cannot grow forever. The property of being an expander means that the graph grows exponentially until about half the vertices have been covered.

\item The diameter of a graph is the maximum distance between any two vertices. The key fact about expander graphs is that their  diameters are logarithmic in $N$.

\end{enumerate}

\section{High temperature expansion}
\label{sctn:high temperature}

Here we consider the high temperature expansion for the sparse model defined from a regular hypergraph with coupling variance $\langle J_A^2\rangle = J^2/kq$.

The partition function of a single disorder realization is
\begin{equation}
    Z = \text{tr} e^{-\beta H}.
\end{equation}
In a high temperature expansion, this is
\begin{equation}
    Z=\text{tr}(1) \left(1 + \frac{\beta^2}{2} \text{tr}(\rho_\infty H^2) + \frac{\beta^4}{4!} \text{tr}(\rho_\infty H^4) + \cdots \right).
\end{equation}
Note that terms with odd powers of $H$ give zero with high probability. The two leading terms are
\begin{equation}
    C_1= \text{tr}(\rho_\infty H^2) = \frac{1}{2^q} \sum_A J_A^2 x_A
\end{equation}
and
\begin{equation}
    C_2 = \text{tr}(\rho_\infty H^4) = \frac{3}{2^{2q}} \sum_{A\neq B} J_A^2 J_B^2 x_A x_B + \frac{1}{2^{2q}} \sum_A J_A^4 x_A  -\frac{2}{2^{2q}} \sum \limits_{A\cap B \in odd} J_A^2 J_B^2 x_A x_B
\end{equation}
The last term coming from commuting different terms in the Hamiltonian. 

From the partition function, the leading terms in the energy before averaging the $J$ couplings are
\bea
E =-C_1 \beta +\frac{\beta^3}{6}(3 C_1^2 -C_2)+\cdots.
\eea
The disorder average of the first term is
\bea
E^{(1)} =- \beta \braket{C_1}_J =-\frac{\beta J^2N}{q2^{2q}},
\eea
which is independent of the graph structure and is the same as in the all-to-all connected SYK model. 
At the next order of $\beta^3$, most terms cancels as expected since the energy is extensive. Its disorder average is
\bea
E^{(3)} =  \frac{\beta^3}{3\cdot 2^{2q}}\left \langle \sum J_A^4 + \sum\limits_{|A\cap B|\in odd}  J_A^2 J_B^2 \right \rangle  =\frac{N\beta ^3 J^4}{3\cdot 2^{2q}} \left ( 1 -\frac{(q-3)}{kq^2} \right ) .
\eea 
We have used the condition that on a regular hypergraph an interaction term connects to $(kq-1)q$ other interaction terms through a single majorana. The probability that two terms share more than one majorana is suppressed by factors of $1/N$.
Therefore, to third order in $\beta$, the disorder averaged energy is
\bea
\frac{E}{N} = -\frac{\beta J^2}{q2^{2q}} + \frac{\beta ^3 J^4}{3\cdot 2^{2q}} \left ( 1 -\frac{(q-3)}{kq^2} \right ) +\cdots.
\eea
This predicts a $k$ dependent correction $(q-3)/kq^2$ to the average energy at order $\beta^3$. While non-zero, this correction is practically quite small, for example, the correction is about $1.5\%$ for $k=4$ and $q=4$.

\section{Disorder fluctuations}
\label{sctn:flctn}

\begin{figure}
\includegraphics[height=0.45\columnwidth, width=0.9\columnwidth]
{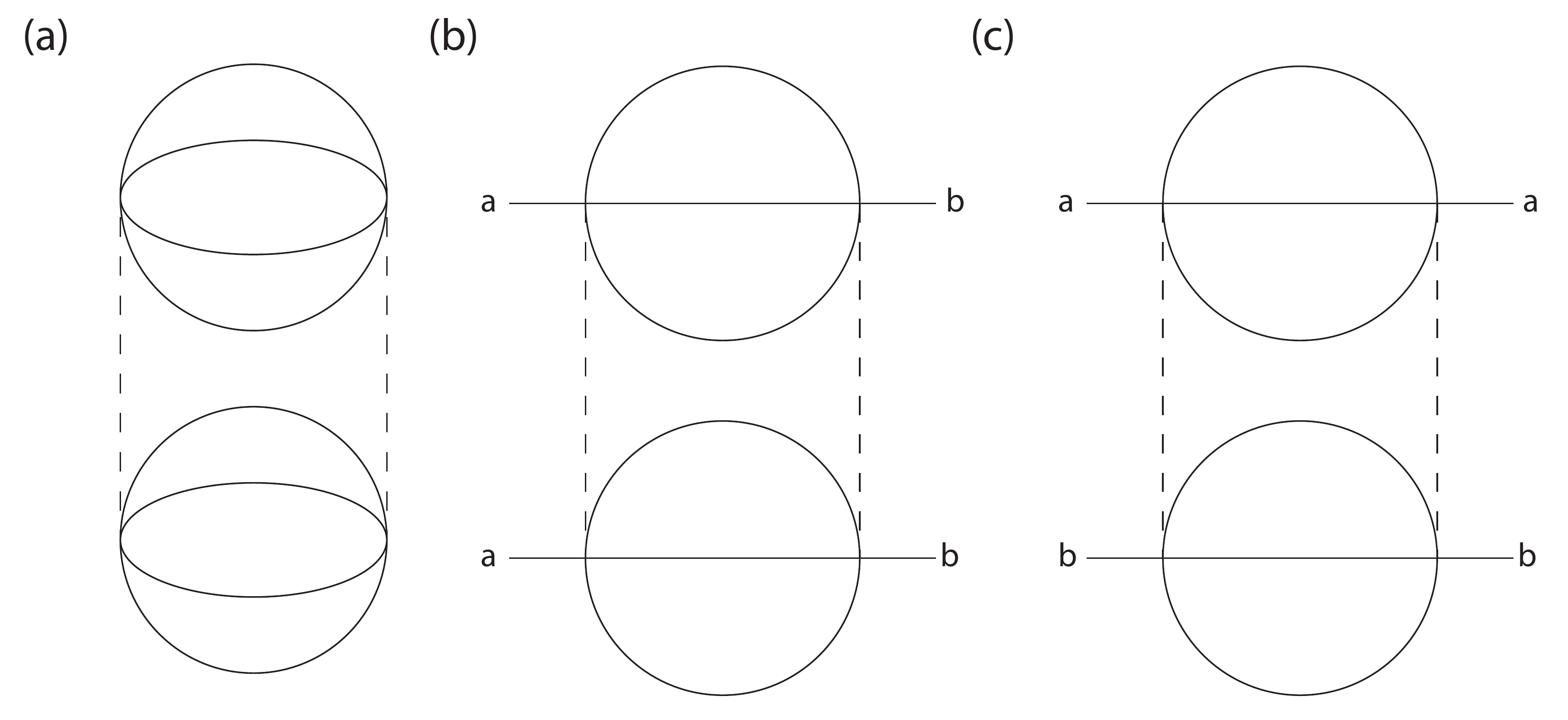}
\caption{(a) The leading Feynman diagram for the replica non-diagonal contribution to the action. (b) The leading Feynman diagram  for the disorder variance of $\braket{\mathcal{T} \chi_a(\tau)\chi_b(0)}$. (c) The leading diagram that contributes to the disorder variance of the on-site Green's function $C^G(a,b)$. The dashed line represents disorder averaging. }
\label{fig:flctn_diagonal}
\end{figure}

This section discusses the sample-to-sample fluctuations of various physical quantities in the sparse SYK model. We first notice that the sparse model is not self-averaging as in the all-to-all case, i.e. $\braket{Z}^m\neq \braket{Z^m}$. The easiest way to see the difference is from the Feynman diagram in Fig.~\ref{fig:flctn_diagonal}(a), which is the leading diagram for the replica non-diagonal contribution to the partition function. In the all-to-all case, the diagram is $\sim J^4/N^{q-2}$. On the other hand, the diagram in the sparse case is $\sim J^4N/kq$. 
Although the difference is obvious from the Feynman diagram, it is non-trivial to study it from the path integral formalism. The replica non-diagonal fluctuations on top of the replica diagonal saddle are independent of the interaction and are zero unless we expand the action to the $q$th order. Another possibility is that there exist replica non-diagonal saddles that dominate over the replica diagonal saddle in certain regimes. If this is the case, then the dominating non-diagonal saddle must be spatially non-uniform, meaning the solution depends on the fermion index, since it was shown that among uniform solutions the replica diagonal saddle point always dominates~\cite{Arefeva2018}. In any case, this suggests that it is difficult to analyze the sample-to-sample fluctuations using the path integral formalism. Therefore we switch to Feynman diagrams for this purpose.

-- \textit{Green's function } We first discuss the disorder fluctuations in the Green's functions. The Green's function $G_{ab}(\tau)=\braket{\mathcal{T}\chi_a(\tau)\chi_b}$ averages to zero for $a\neq b$. The leading diagram for its disorder variance is given in Fig.~\ref{fig:flctn_diagonal}(b). In the fully-connected SYK model, this diagram scales as $1/N^{q-1}$. In the sparse case, it is only non-vanishing if there are two hyperedges having the same vertices except $a$ and $b$. This requires there to be small loops in the sparse hpergraph, the probability of which is suppressed by $1/N^{q-1}$. Therefore the disorder variance of $G_{ab}$ has the same scaling behavior in the sparse model and fully connected model for $a\neq b$.

In contrast, the disorder fluctuation in on-site Green's function $G_{aa}$ significantly increases due to the sparseness of the graph structure. In the fully connected model, from the Feynman diagram in Fig.~\ref{fig:flctn_diagonal}(c) we can see that $C^G(a,b)=\braket{G_{aa} G_{bb}}_J - \braket{G_{aa}}_J\braket{G_{bb}}_J$ scales as $1/N^{q-1}$ if $a=b$ and $1/N^q$ if $a\neq b$, all suppressed by powers of $1/N$. Therefore the Green's functions are self-averaging in the fully-connected SYK model.  However, the disorder fluctuations in the sparse case are not suppressed by $1/N$ but $1/k$ and depend on the distance between $a$ and $b$ on the hypergraph. Let us first consider the case $a=b$. The leading contribution is also from the diagram in Fig.~\ref{fig:flctn_diagonal}(c),  suggesting that $C^G(a,a)$ scales as $(1/kq)^2 kq =1/kq$, where $(1/kq)^2$ comes from the coupling averaging and $kq$ is the number of hyperedges (terms in the Hamiltonian) containing $a$. 

\begin{figure}
\includegraphics[height=0.5\columnwidth, width=0.5\columnwidth]
{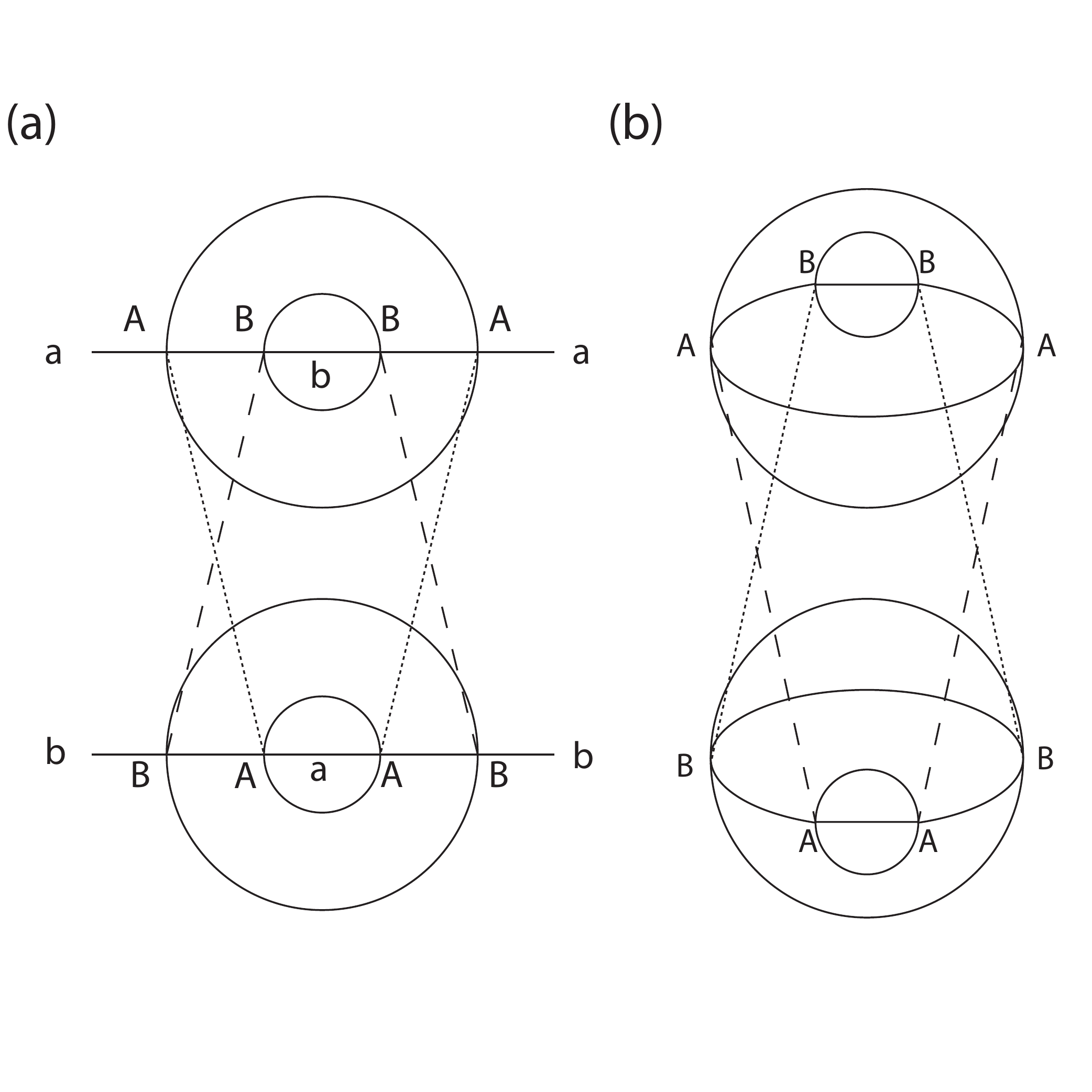}
\caption{(a) The leading Feynman diagram in $C^G(a,b)$ when $a$ and $b$ are nearest neighboring vertices. (b) The leading Feynman diagram in $C^E(A,B)$ when $A$ and $B$ are nearest neighboring hyperedges. The dashed lines represents disorder average, and dashed lines with different spacing are associated with different coupling terms.}
\label{fig:flctn_cross}
\end{figure}

Now, when $a$ and $b$ are not the same but in the same hyperedge, the diagram indicates $C^G(a,b)\sim 1/(kq)^2$ since the probability of more than one hyperedges containing both $a$ and $b$ are suppressed by large $N$. 
When $a$ and $b$ are in two neighboring hyperedges, i.e., $\text{dist}(a,b)=2$, the leading Feynman diagram for the disorder fluctuation is sketched in Fig.~\ref{fig:flctn_cross}(a). The diagram suggests that $C^G (a,b)\sim 1/(kq)^4$ in this case. As $a$ and $b$ are separated further, the leading Feynman diagram includes interaction couplings $J_C$ for all the hyperedges $C$ in the shortest path between $a$ and $b$, with each factor of $J_C^4$ contributing a factor of $(kq)^{-2}$ after disorder averaging.

The above analysis suggests that the onsite Green's function is not self-averaging in the sparse case. In addition, $C^G(a,b)\sim (kq)^{-2\,\text{dist}(a,b)}$, so that the correlation between Green's functions on different vertices decays exponentially with distance on the hypergraph. In principle, we should take account of different paths connecting $a$ and $b$ on the graph. However, the local structure of a sparse hypergraph is tree-like, so for vertices a finite distance apart, there is typically only one short path between them. With these tools, we can now analyze the disorder fluctuation in the averaged Green's function $\sum G_a/N$. The fluctuations are given by $\frac{1}{N^2}\sum\limits_{a,b} C^G(a,b)$.

Given a vertex $a$, there are roughly $(kq^2)^d$ vertices a distance $d$ away. Using $C^G(a,b)\sim (kq)^{-2\,\text{dist}(a,b)}$, we get
\bea
\sum\limits_b C^G(a,b)\sim \sum\limits_d k^{-d} \sim \text{constant}.
\eea
Note that we have assumed that $k>1$. As a result, the disorder variance of the site-averaged Green's function scales like $1/N$. Therefore the site-averaged Green's function is self-averaging even in the sparse model. 

--\textit{Energy } Now we turn to the thermal energy of the sparse model. We have gained some intuition from the sample-to-sample fluctuations in the Green's function and its relation to the graph structure. It turns out that the disorder fluctuation in the energy behaves in a similar way.

The total energy is a sum of the local energies $E_A$ defined on the hyperedges and we define the connected correlation function
\bea
C^E (A, B) = \braket{E_A E_B}_J -\braket{E_A}_J \braket{E_B}_J.
\eea
The energy variance is
\bea
\text{var}(E) = \sum\limits_A C^E(A,A) + \sum\limits_{A\neq B} C^E(A,B)
\eea
In the fully connected model, the main contribution is the from the first term $C^E(A,A)$ which scales $1/N^{(2q-2)}$ based on Fig.~\ref{fig:flctn_diagonal}. There are $N^q$ of them and the total energy variance scales $1/N^{q-2}$, self-averaging for $q>2$. 

In the sparse model, $C^E(A,A)$ scales $1/(kq)^2$ and therefore the first term is extensive. The second term depends on how fast the correlation decays on the hypergraph. In the case that $A$ and $B$ are nearest neighboring, i.e. connected by a vertex, the leading diagram for $C^E(A,B)$ is shown in Fig.~\ref{fig:flctn_cross}(b). The Feynman diagram scales as $1/(kq)^4$. In general $C^E(A,B)$ decays as $(kq)^{-2\, \text{dist}(A,B)}$. Given $A$, there are $(kq^2)^d$ hyperedges a distance $d$ away. Based on the same argument as for the Green's function, the variance of energy is also extensive. Therefore the energy density is self-averaging but the total energy is not.

We can also explicitly calculate the disorder fluctuation in the energy using the results from the high temperature expansion. Recall that, before the disorder averaging, the first term in the high temperature expansion of the energy is
\bea
E^{(1)} = -\frac{\beta}{2^q} \sum\limits_A J_A^2 x_A.
\eea
Clearly, $\text{var}(E^{(1)})$ is extensive. The next order is
\bea
E^{(3)} = \frac{\beta^3}{3\cdot 2^{2q}}\sum\limits_A\left(  J_A^4 x_A + \sum\limits_{B,|B\cap A|=1}  J_A^2 J_B^2 x_A x_B \right )
\eea
Since each term in the sum only correlates with finite number of other terms, its variance is again extensive. 

\begin{figure}
\includegraphics[height=0.5\columnwidth, width=0.5\columnwidth]
{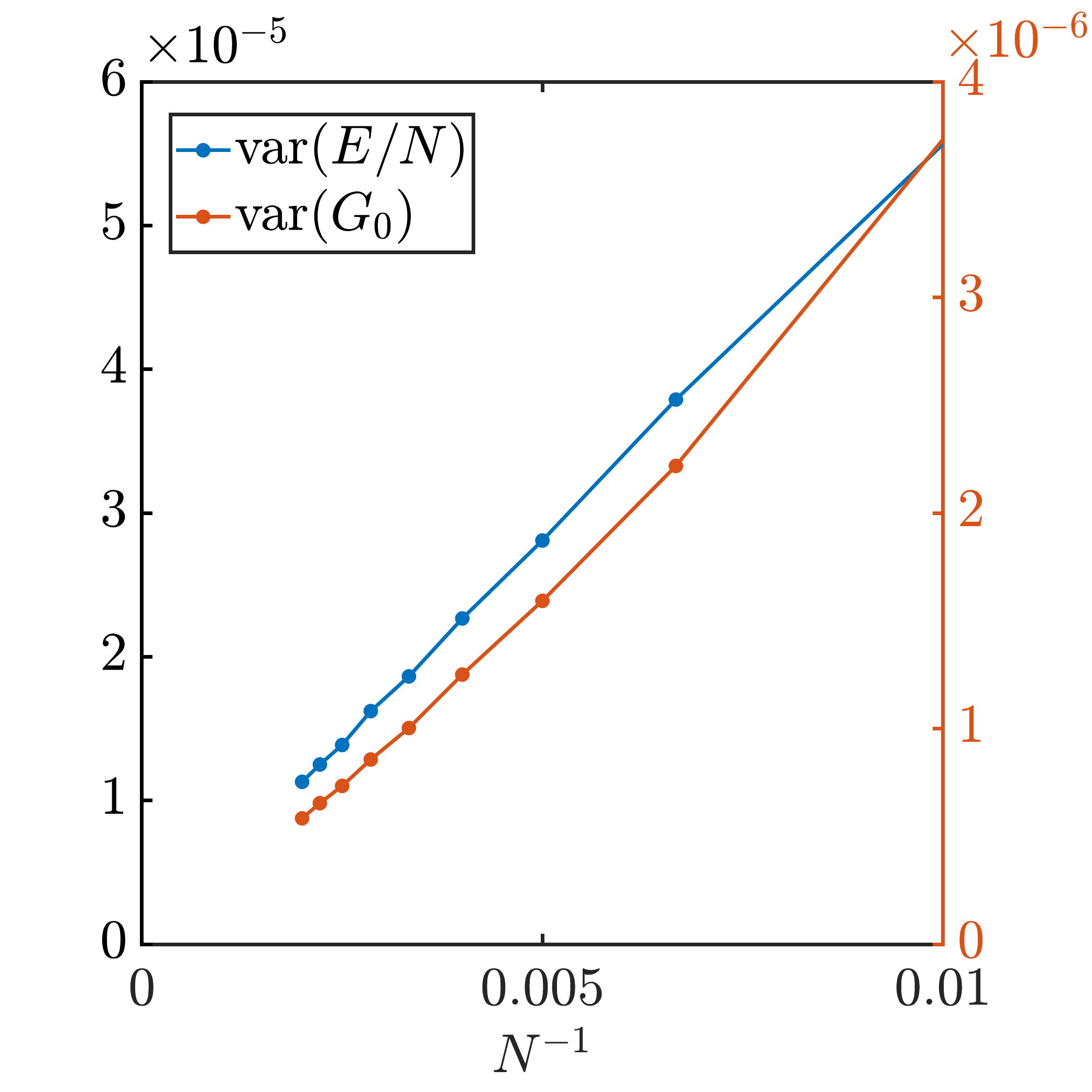}
\caption{The disorder variance for the energy density and site-averaged Green's function in the SYK$_2$ model. Both quantities decreases as $1/N$, showing that the energy density and the site-averaged Green's function are self-averaging.}
\label{fig:flctn}
\end{figure}

Next we verify numerically the scaling of the disorder fluctuations in the case of $q=2$, where the Green's function and energy can be calculated for large system size because the model is non-interacting. 
The Hamiltonian is 
\begin{equation}
    H = i \sum_{ab} J_{ab} x_{ab} \chi_a \chi_b.
\end{equation}
The couplings $J_{ab}$ are anti-symmetric, while the $x_{ab}\in \{0,1\}$ determine the interaction graph. Because the coupling matrix is anti-symmetric, eigenvalues come in plus/minus pairs. In terms of the eigenvalue pairs, $\pm \epsilon_i$, the energy and Green's function at temperature $T=\beta^{-1}$ is
\bea
    E = \sum_{i=1}^{N/2} - \frac{\epsilon_i}{2} \tanh \frac{\beta \epsilon_i}{2}, \quad G^{ab}(\tau) =\frac{1}{2}\sum\limits_{i=1}^N u_{ai} \left (e^{-\frac{(\beta-2\tau)\epsilon_i}{2}}\cosh^{-1} \frac{\beta\epsilon_i}{2}\right ) u^*_{ib}
\eea
In Fig.~\ref{fig:flctn} we present the disorder variance of the energy density $E/N$ and the site-averaged Green's function $G_0(\beta/2) =\frac{1}{N} \sum G_{aa}(\beta/2)$ for $\beta=100/J$ and 30000 disorder couplings and random prunings fixing $k=4$. They both decrease as $1/N$, which agrees with the general argument above. 


To conclude this section, in the sparse model, neither Green's function and energy is self-averaging, but the site-averaged Green's function and the energy density are self-averaging for large enough but $N$-independent $k$.


\section{Simulating all-to-all model with product formulas}
\label{sctn:all2all}
In this section, we analyze the complexity of the product-formula algorithm for simulating the SYK model with all-to-all couplings. Recall that the Hamiltonian has the form
\begin{equation}
    H_{\text{full}} = \frac{i^{q/2}}{q!} \sum_{a_1 \neq \cdots \neq a_q} J_{a_1\cdots a_q} \chi_{a_1} \cdots \chi_{a_q},
\end{equation}
where $\chi_a$ are Majorana fermion operators and
\begin{equation}
    \sqrt{\langle J_{a_1 \cdots a_q}^2\rangle_J} = \frac{(q-1)!^{\frac{1}{2}} J}{N^{\frac{q-1}{2}}}.
\end{equation}
Here, we use the convention that $\chi_a^2=I/2$. To analyze the performance of product formulas, we need to estimate the norm of all nested commutators of the terms from the Hamiltonian $\alpha_p=\sum_{i_1,\ldots,i_{p+1}}||W_{i_1\cdots i_{p+1}}||$. Here, the nested commutators are defined recursively through
\begin{equation}
    W_{i_1\cdots i_{l+1}}:=\left[H_{i_{l+1}},W_{i_1\cdots i_{l}}\right],
\end{equation}
where $H_{i_{l+1}}$ is an arbitrary Hamiltonian term, i.e., a string of $q$ Majorana fermion operators.

Given two Majorana strings with even length, their commutator is zero unless the supports overlap. In this case, the commutator either gives zero, or a new Majorana string whose length is the sum of the length of the original strings. Using this property, we deduce that $W_{i_1\cdots i_{l+1}}$ is either zero or a Majorana string, which has length $(l+1)q-2l$ and overlaps with $((l+1)q-2l)N^{q-1}/(q-1)!$ terms in the Hamiltonian that contribute to $\alpha_p$. Therefore, we have
\begin{equation}
\begin{aligned}
    \alpha_p&=\mathcal{O}\left(\prod_{i=1}^{p}(iq-2i+2)\frac{N^{p(q-1)+q}}{(q-1)!^{p}q!}
    2^{-\frac{(p+1)q-2p}{2}}\frac{(q-1)!^{\frac{p+1}{2}}J^{p+1}}{N^{\frac{(p+1)(q-1)}{2}}}\right)\\
    &=\mathcal{O}\left(\frac{p! q^p N^{p(q-1)+q} (q-1)!^{\frac{p+1}{2}}J^{p+1}}{(q-1)!^{p}q! 2^{\frac{(p+1)q-2p}{2}} N^{\frac{(p+1)(q-1)}{2}}}\right),
\end{aligned}
\end{equation}
which simplifies to
\begin{equation}
    \alpha_p=\mathcal{O}\left(\frac{e^{\frac{(p+1)q}{2}} N^{\frac{pq-p+q+1}{2}}J^{p+1} }{ 2^{\frac{(p+1)q}{2}} q^{\frac{pq}{2}-\frac{5}{4}p+\frac{q}{2}+\frac{3}{4}} }\right)
\end{equation}
through Stirling's formula. As a result, we need
\begin{equation}
    n=\mathcal{O}\left(e^{(1+\delta)q/2}2^{-(1+\delta)q/2} N^{q-1+(q+1)\delta} q^{\frac{q}{2}-\frac{5}{4}+(\frac{q}{2}+\frac{3}{4})\delta} (Jt)^{1+\delta}\epsilon^{-\delta}\right)
\end{equation}
steps to simulate the evolution for time $t$ with accuracy $\epsilon$.

In each step, we need to implement
\begin{equation}
    \frac{N^q}{q!}=\mathcal{O}\left(\frac{N^q e^q}{q^{q+\frac{1}{2}}}\right)
\end{equation}
elementary exponentials. Using the Bravyi-Kitaev encoding, the simulation has a total cost of
\begin{equation}
    \mathcal{O}\left(e^{(3+\delta)q/2}2^{-(1+\delta)q/2} N^{2q-1+(q+1)\delta} q^{(\frac{q}{2}+\frac{3}{4})(-1+\delta)} (Jt)^{1+\delta}\epsilon^{-\delta}
    \log N\right).
\end{equation}

\end{appendices}

\end{document}